\documentclass[twocolumn,aps,prd,superscriptaddress,longbibliography]{revtex4-2}
\usepackage{amsmath}
\usepackage{amssymb}
\usepackage{graphicx}
\usepackage{color}
\usepackage[colorlinks=true,linkcolor=blue,citecolor=blue]{hyperref}
\usepackage{tabularx}
\usepackage{float}
\usepackage{array}
\usepackage{multirow}
\usepackage{arydshln}
\usepackage{booktabs}

\newcommand{\be}{\begin{equation}}
\newcommand{\ee}{\end{equation}}
\newcommand{\bea}{\begin{eqnarray}}
\newcommand{\eea}{\end{eqnarray}}
\newcommand{\bse}{\begin{subequations}}
\newcommand{\ese}{\end{subequations}}

\newcommand{\bfa}{\mathbf{a}}
\newcommand{\bfk}{\mathbf{k}}

\newcommand{\bfrr}{\mathbf{R}}
\newcommand{\bfgg}{\mathbf{G}}
\newcommand{\bfqq}{\mathbf{Q}}
\newcommand{\bfss}{\mathbf{S}}

\definecolor{go_green}{rgb}{0.13, 0.55, 0.13}

\begin{document}

\author{Farhan Islam}
\affiliation{Ames National Laboratory, Iowa State University, Ames, Iowa 50011, USA}
\affiliation{Department of Physics and Astronomy, Iowa State University, Ames, Iowa 50011, USA}
\author{Tha\'{i}s V. Trevisan}
\affiliation{Ames National Laboratory, Iowa State University, Ames, Iowa 50011, USA}
\affiliation{Department of Physics and Astronomy, Iowa State University, Ames, Iowa 50011, USA}
\author{Thomas Heitmann}
\affiliation{The Missouri Research Reactor and Department of Physics and Astronomy, University of Missouri, Columbia, Missouri 65211, USA}
\author{Santanu Pakhira}
\author{Simon~X.~M.~Riberolles}
\author{N. S. Sangeetha}
\altaffiliation[Present Address: ]{Institute for Experimental Physics IV, Ruhr University Bochum, 44801 Bochum, Germany}
\affiliation{Ames National Laboratory, Iowa State University, Ames, Iowa 50011, USA}

\author{David C. Johnston}
\author{Peter P. Orth}
\author{David Vaknin}
\email{vaknin@ameslab.gov}
\affiliation{Ames National Laboratory, Iowa State University, Ames, Iowa 50011, USA}
\affiliation{Department of Physics and Astronomy, Iowa State University, Ames, Iowa 50011, USA}

\title{Frustrated Magnetic Cycloidal Structure and Emergent Potts Nematicity in CaMn$_2$P$_2$}
\date{\today}

\begin{abstract}
We report neutron-diffraction results on single-crystal CaMn$_2$P$_2$ containing corrugated Mn honeycomb layers and determine its ground-state magnetic structure. The diffraction patterns consist of prominent (1/6, 1/6, $L$) reciprocal lattice unit (r.l.u.; $L$ = integer) magnetic Bragg reflections, whose temperature-dependent intensities are consistent with a first-order antiferromagnetic phase transition at the N\'eel temperature $T_{\rm N} = 70(1)$ K. Our analysis of the diffraction patterns reveals an in-plane $6\times6$ magnetic unit cell with ordered spins that in the principal-axis directions rotate by 60-degree steps between nearest neighbors on each sublattice that forms the honeycomb structure, consistent with the $P_Ac$ magnetic space group. We find that a few other magnetic subgroup symmetries ($P_A2/c$, $P_C2/m$, $P_S\bar{1}, P_C2, P_Cm, P_S1$) of the paramagnetic $P\bar{3}m11^\prime$ crystal symmetry are consistent with the observed diffraction pattern. We relate our findings to frustrated $J_1$-$J_2$-$J_3$ Heisenberg honeycomb antiferromagnets with single-ion anisotropy and the emergence of Potts nematicity.
\end{abstract}
\maketitle

\section{Introduction}
Magnetic materials with local moments arranged on a honeycomb lattice are known to exhibit a variety of complex magnetic states in the presence of frustrated spin exchange interactions. Recent examples are the honeycomb iridates~\cite{Chaloupka2010,Trebst2022}, the nickelate Ni$_2$Mo$_3$O$_8$~\cite{Morey2019}, transition metal oxides ${\text{InCu}}_{2/3}{\text{V}}_{1/3}{\text{O}}_{3}$~\cite{Yehia2010,Iakovleva2019}, Bi$_3$Mn$_4$O$_{12}$(NO$_3$)~\cite{Smirnova2009} and verdazyl-based salts~\cite{Miyamoto2019}. Often the complex behavior of these systems can be rationalized using quantum spin models such as the Kitaev-Heisenberg honeycomb model~\cite{Kitaev2006,Baskaran2008,Reuther2011,Price2013} or the $J_1$-$J_2$-$J_3$ Heisenberg honeycomb model~\cite{Rastelli1979,Katsura1986,Fouet2001,Mulder2010,Albuquerque2011,Reuther2011a,Oitmaa2011,Clark2011,Bishop2012,Bishop2013b,Bishop2015,Sahoo2020}. The former exhibits various complex magnetically-ordered phases and a quantum-spin-liquid ground state when Kitaev interactions are dominant and the local moments have low spin $S =1/2,1, 3/2$~\cite{Kitaev2006,Dong2020,Jin2022}. The latter hosts different collinear and non-collinear magnetic states, including complex spirals, already in the classical limit, and its phase diagram also includes magnetically-disordered regions with intriguing valence-bond crystal correlations for $S=1/2$~\cite{Albuquerque2011}. Specifically, for $J_3=0$, the classical $J_1$-$J_2$ Heisenberg honeycomb antiferromagnet exhibits a N\'eel-ordered ground state for $J_2 < J_1/6$ and  degenerate single-$Q$ spiral states for $J_2 > J_1/6$~\cite{Rastelli1979,Katsura1986,Fouet2001,Mulder2010}. Nonzero $J_3$ or, alternatively, quantum and thermal fluctuations~\cite{Mulder2010} lift this continuous degeneracy and select six symmetry-related wavevectors out of the degenerate manifold.
\begin{figure}[t!]
\centering
\includegraphics[width=2.8 in]{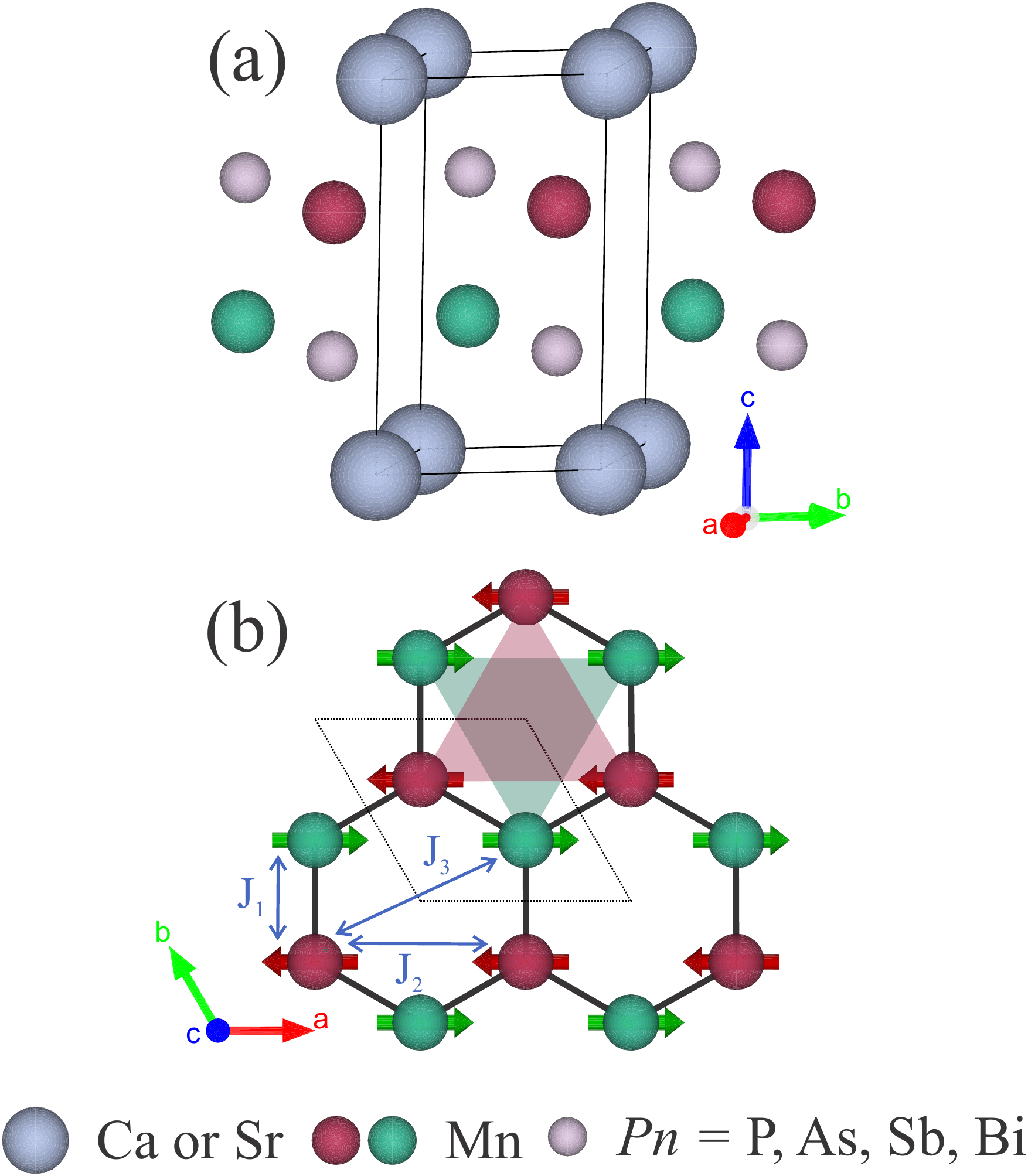}
\caption{ (a) Chemical structure of AMn$_2$Pn$_2$ (A = Sr, Ca; Pn = P, As, Sb, Bi) showing the Mn trigonal bilayer without intervening elements. (b) Projection of the two trigonal Mn sublattices onto the $ab$-plane, shown with red and green shades. The A and B layers are stacked with two atoms per unit cell (the dotted rhombus shows the basal unit cell). The Mn bilayer forms a corrugated honeycomb lattice, where the nearest-neighbor (NN) interactions ($J_1$) and the next-nearest-neighbor (NNN) interactions ($J_2$) are indicated. The magnetic structure shown is typical for the AMn$_2$Pn$_2$ compounds with $Pn=$ As, Sb, Bi, for which the first-neighbor interactions are dominant and antiferromagnetic: $J_{1} \gg J_{2}, J_3$. In contrast, here we report that CaMn$_2$P$_2$ exhibits a different magnetic structure that emerges mainly due to frustrated couplings $J_{1}/2 \approx J_2.$ (Although implied in the figure, SrMn$_2$Bi$_2$ has not yet been synthesized or discussed in the literature.)}
\label{Fig:Structure}
\end{figure}

Another rich experimental platform for frustrated honeycomb magnets consist of the trigonal compounds CaMn$_2$P$_2$, CaMn$_2$As$_2$, CaMn$_2$Sb$_2$, CaMn$_2$Bi$_2$, SrMn$_2$P$_2$, SrMn$_2$As$_2$, SrMn$_2$As$_2$ and with space group $P\bar{3}m1$ (no. 164) \cite{Sangeetha2021,Sangeetha2016,Simonson2012a,Sangeetha2018,Das2017,Bridges2009,Gibson2015} and associated point group $D_{3d}$. As shown in Fig.~\ref{Fig:Structure}, these systems contain the transition-metal element Mn in a corrugated honeycomb structure, which is formed by two adjacent trigonal layers (or sublattices) that are stacked in an A-B type fashion. The Mn atoms occupy Wyckhoff positions $2d$ with site symmetry $3m$. There are two Mn atoms per unit cell, which form the A, B sublattice sites of the honeycomb lattice.  The transition-metal bilayer magnetic moments have no intervening binding atoms, as shown in Fig.~\ref{Fig:Structure}(a), so that the major magnetic coupling  between nearest neighbors is likely a direct Mn-Mn coupling, and couplings among next-nearest neighbors (NNN) are likely due to Mn-$Pn$-Mn superexchange. Neutron diffraction measurements of Mn compounds with $Pn =$ As, Sb, or Bi have revealed a simple N\'eel-type magnetic structure in SrMn$_2$As$_2$, CaMn$_2$Sb$_2$, and CaMn$_2$Bi$_2$~\cite{Das2017,Simonson2012a,Gibson2015,Ratcliff2009}. This N\'eel magnetic structure is shown schematically in Fig.~\ref{Fig:Structure}(b). For CaMn$_2$Sb$_2$, it has been suggested that the moments are slightly canted towards the $c$-axis~\cite{Bridges2009}. These observations are consistent with a dominant NN interaction $J_1 \gg J_2$ for these materials.


It has recently been concluded that the superexchange within an Mn-$Pn$-Mn moiety increases as the atomic number of $Pn$ is reduced, thereby increasing the magnetic frustration in the system. Thus, NNN are expected to be stronger for $Pn =$ P than for $Pn =$ Bi, for similar bond configurations~\cite{Islam2020a}. We thus expect CaMn$_2$P$_2$ to experience a sizable NNN coupling $J_2$ and thus substantial magnetic frustration, which is one of the main motivations for this work. 


Here, we report neutron-diffraction results on single-crystals of CaMn$_2$P$_2$, and determine its ground-state magnetic structure. 
Recent $^{31}$P NMR measurements \cite{Sangeetha2021} indicate that the magnetic structure of CaMn$_2$P$_2$ is commensurate with the lattice. This is in contrast to SrMn$_2$P$_2$ that was found to experience an  incommensurate magnetic order~\cite{Sangeetha2021}. These observations are consistent with neutron-diffraction measurements of SrMn$_2$P$_2$ that indicate a complex and so far undetermined magnetic structure~\cite{Brock1994}. Interestingly, CaMn$_2$P$_2$ and SrMn$_2$P$_2$ have recently been reported to undergo an unusual first-order antiferromagnetic (AFM) transition at $T_{\rm N} =70(3)$ and 53(1)~K, respectively~\cite{Sangeetha2021}. By contrast, the isostructural CaMn$_2$As$_2$ and SrMn$_2$As$_2$ compounds undergo second-order AFM transitions~\cite{Sangeetha2016}. Below, we relate the observed first-order magnetic transition in CaMn$_2$P$_2$ with its more complex spiral magnetic order that breaks threefold rotation symmetry and promotes the emergence of a Potts-nematic order parameter~\cite{Mulder2010,Little2020,Fernandes2019}. 

We note that  $A$Mn$_2Pn_2$ ($A=$ Ca or Sr and $Pn =$ P, As, Sb) compounds display strong two-dimensional (2D) magnetic fluctuations as manifested in magnetic susceptibility ($\chi $) measurements that do not show Curie-Weiss behavior at temperatures much higher than $T_{\rm N}$ \cite{Das2017,Ratcliff2009,Bridges2009,Gibson2015,Sangeetha2021}. In addition, the $\chi (T)$ with applied magnetic field along the $ab$-plane for all these compounds hardly shows any anomaly at $T_{\rm N}$. This 2D behavior is also manifested in the magnetic order parameter in neutron-diffraction measurements of SrMn$_2$As$_2$~\cite{Das2017}. These characteristics indicate that the dominant in-plane NN coupling $J_1$ is AFM and is likely much larger than the interlayer couplings between honeycomb planes, leading to sizable 2D AFM correlations above $T_{\rm N}$.  Interestingly, inelastic neutron-scattering measurements that were analyzed using spin-wave theory for the $J_1-J_2$ Heisenberg model determined a ratio of  $J_2/J_1 \approx 1/6$ for CaMn$_2$Sb$_2$. This places the system in proximity to a tricritical point that separates a N\'eel ordered phase and two different spiral magnetic phases~\cite{Fouet2001,Oitmaa2011,Mcnally2015}.

\section{\label{Sec:ExpDet} Experimental Details and Methods}
\begin{figure}
\centering
\includegraphics[width=0.95\linewidth]{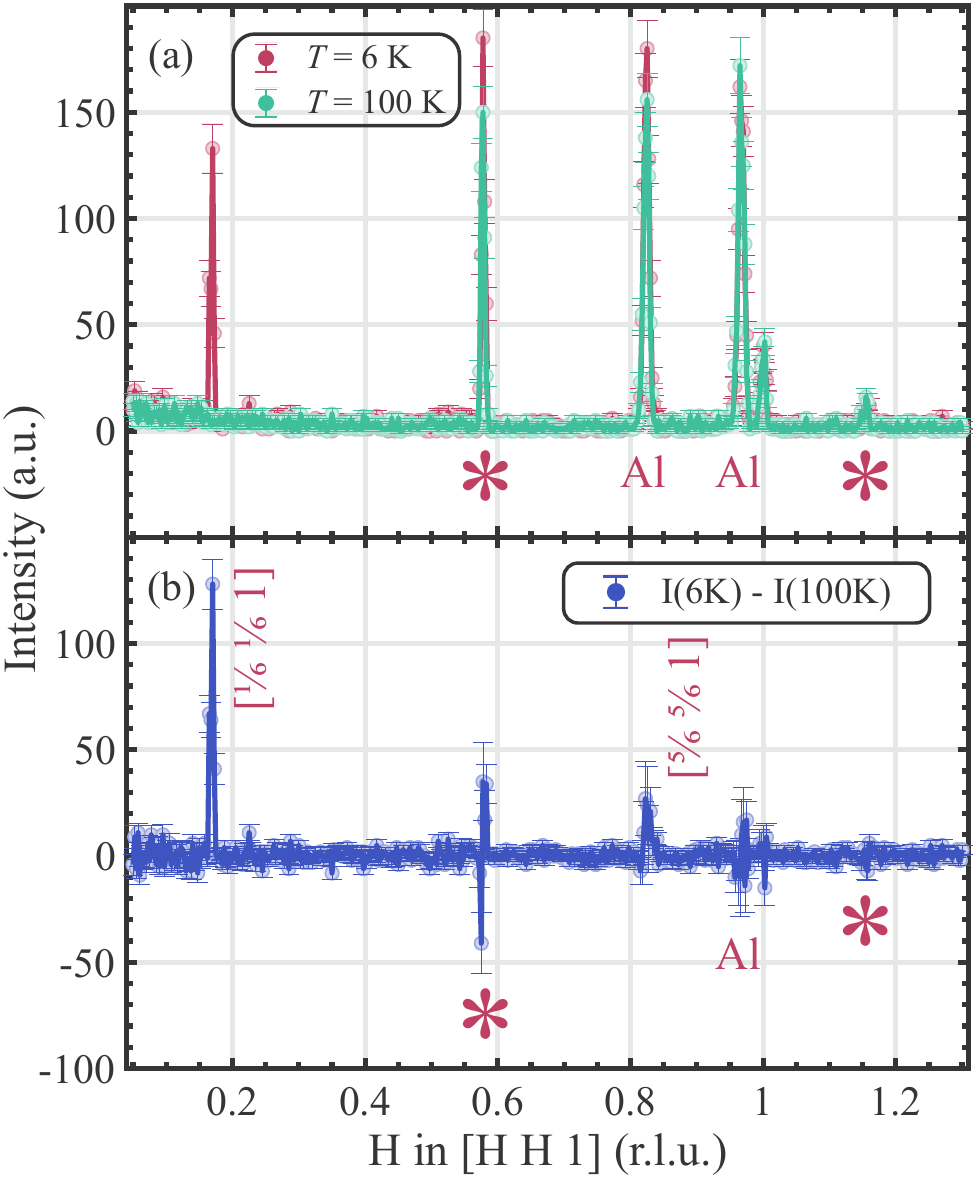}
\caption{(a) Diffraction patterns along $(H,H,1)$ at $T=6$ and 100~K showing the emergence of a prominent peak at $H=1/6$ r.l.u. (b) The difference between the $(H,H,1)$ patterns at 6 K and 100 K showing that the observed magnetic Bragg reflections in this direction are $(\eta,\eta,1)$ and $(1-\eta,1-\eta,1)$, where $\eta=1/6$. Al peaks (originating from the sample holder) in the difference pattern show both positive and negative signals due to the thermal shift in peak positions. The peaks with asterisks originate from a twin of CaMn$_2$P$_2$ oriented in a different direction.}
\label{Fig:hh1}
\end{figure}

Single crystals of CaMn$_2$P$_2$ were grown in Sn flux, as described previously \cite{Sangeetha2021}, and the crystal used in this study is from the same growth batch. Single-crystal neutron-diffraction experiments were performed in zero applied magnetic field using the TRIAX triple-axis spectrometer at the University of Missouri Research Reactor (MURR). An incident neutron beam of energy  14.7 meV was directed at the sample using a pyrolytic-graphite (PG) monochromator. A PG analyzer was used to reduce the background. Shorter neutron wavelengths were removed from the primary-beam using PG filters placed before the monochromator and in between the sample and analyzer. Beam divergence was limited using collimators before the monochromator; between the monochromator and sample; sample and analyzer; and analyzer and detector of $60^\prime-60^\prime-40^\prime-40^\prime$, respectively. A 40~mg CaMn$_2$P$_2$ crystal was mounted on the cold tip of an Advanced Research Systems closed-cycle refrigerator with a base temperature of approximately 5 K. The crystal was mounted in the $(H,0,L)$ and $(H,H,L)$  scattering planes. We measured the lattice parameters to be $a=4.096(1)$ and $c = 6.848(2)$~{\AA} at base temperature. We also note that our sample consists of at least two twins that are disoriented with respect to each other, as indicated in Fig. \ref{Fig:hh1}. Our diffraction patterns here and below also show Bragg reflections from the polycrystalline Al sample holder.

\section{\label{Sec:Neutron} Experimental Results and Analysis}

\begin{figure}[!h]
\centering
\includegraphics[width=0.95\linewidth]{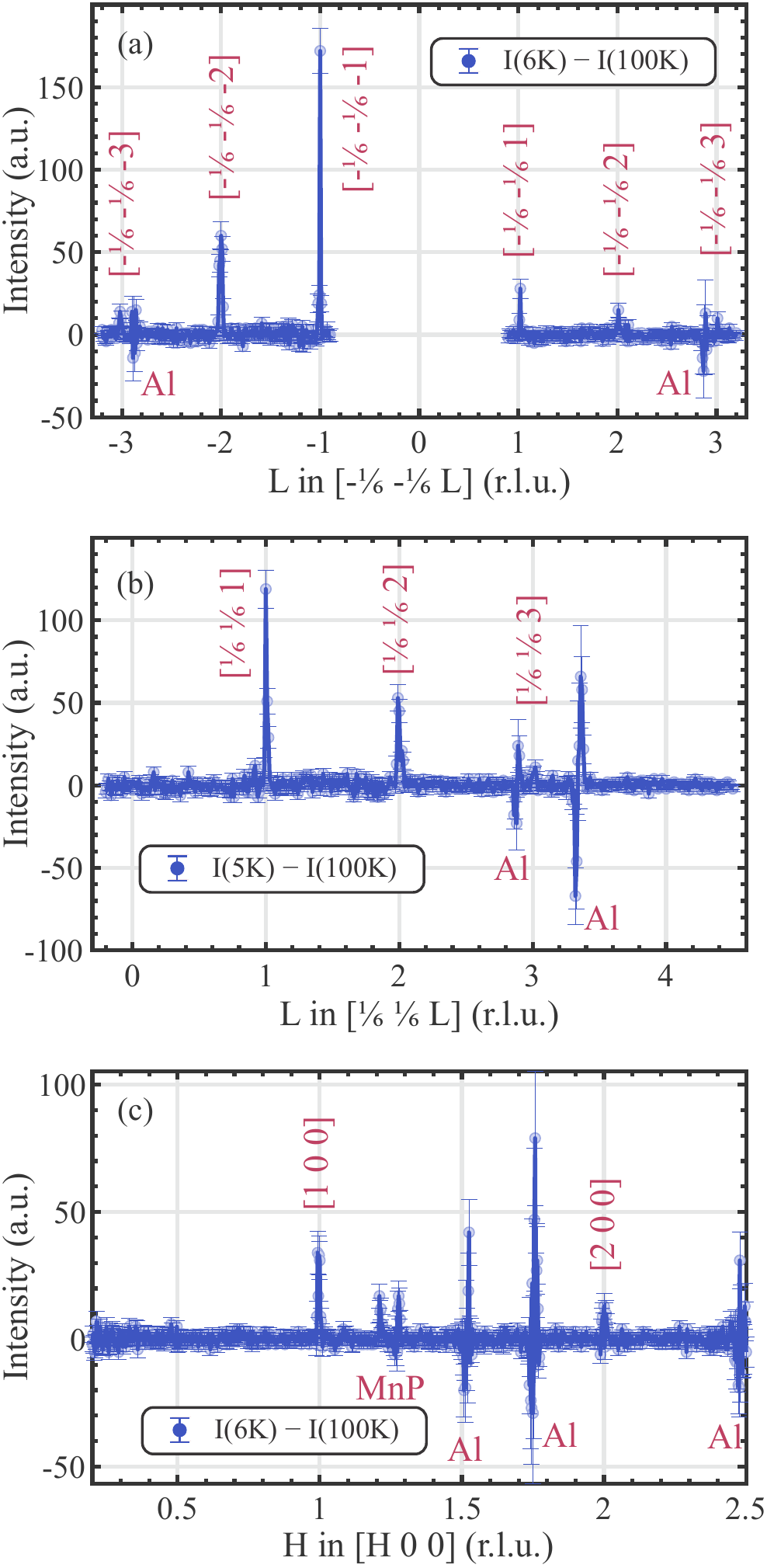}
\caption{Difference between scans at low and high temperature, as indicated, along (a) $(-1/6,-1/6, L)$ showing peaks at integer values of $L$ between $-3$ and $3$ (except for $L=0$); (b) $(1/6,1/6, L)$ showing peaks at the integer values of $L$ between 1 and 3 (scans with negative $L$ were not accessible due to the experimental setup); and (c) $(H,0,0)$ showing a weak peak at the nuclear $(1,0,0)$ position and possibly another at $(2,0,0)$. Signals from the Al sample holder are marked on the figures. As indicated in (c), a minute inclusion of ferromagnetic MnP crystals gives rise to weak peaks.}
\label{Fig:diffraction2}
\end{figure}
\subsection{Experimental results}
Diffraction scans along the $(H,H,1)$ direction at $T=6$ and 100 K in Fig.~\ref{Fig:hh1}(a) show the emergence of a  prominent  peak at $H=1/6$ r.l.u. (reciprocal lattice units) at low temperatures.
As shown in Fig.~\ref{Fig:hh1}(b), the difference between these scans at 6 and 100~K displays magnetic Bragg peaks at $(\eta,\eta,1)$ and $(1-\eta,1-\eta,1)$, where $\eta=1/6$.
Figure \ref{Fig:diffraction2}(a) shows the difference between scans at 6 and 100 K along $(-\eta,-\eta, L)$, indicating magnetic Bragg peaks at $L$ = $-3, -2, -1, 1, 2,$ and $3$. Figure \ref{Fig:diffraction2}(b) shows similar observations of magnetic Bragg peaks at $L$ = 1, 2, and 3 in the direction of $(\eta,\eta, L)$. Scans along $(H,H,0)$ do not show any newly-emerging peaks at low temperatures (not shown).   
Figure \ref{Fig:diffraction2}(c) shows the difference of scans along $(H,0,0)$ at 6 and 100 K with a weak peak at the nuclear $(1,0,0)$  reflection and possibly another very weak one  at $(2,0,0)$ \cite{SM-CaMn2P2}. The other signals that have a negative intensity originate from the Al sample holder. Also, magnetic peaks from a small amount of MnP in the crystal are present in the scan, as indicated.
The temperature dependence of $(1,0,0)$ does not exhibit a transition at $T_{\rm N}$. This implies that that splitting is not significantly related to the observed magnetic structure.
\begin{figure}[h]
\centering
\includegraphics[width=0.95\linewidth]{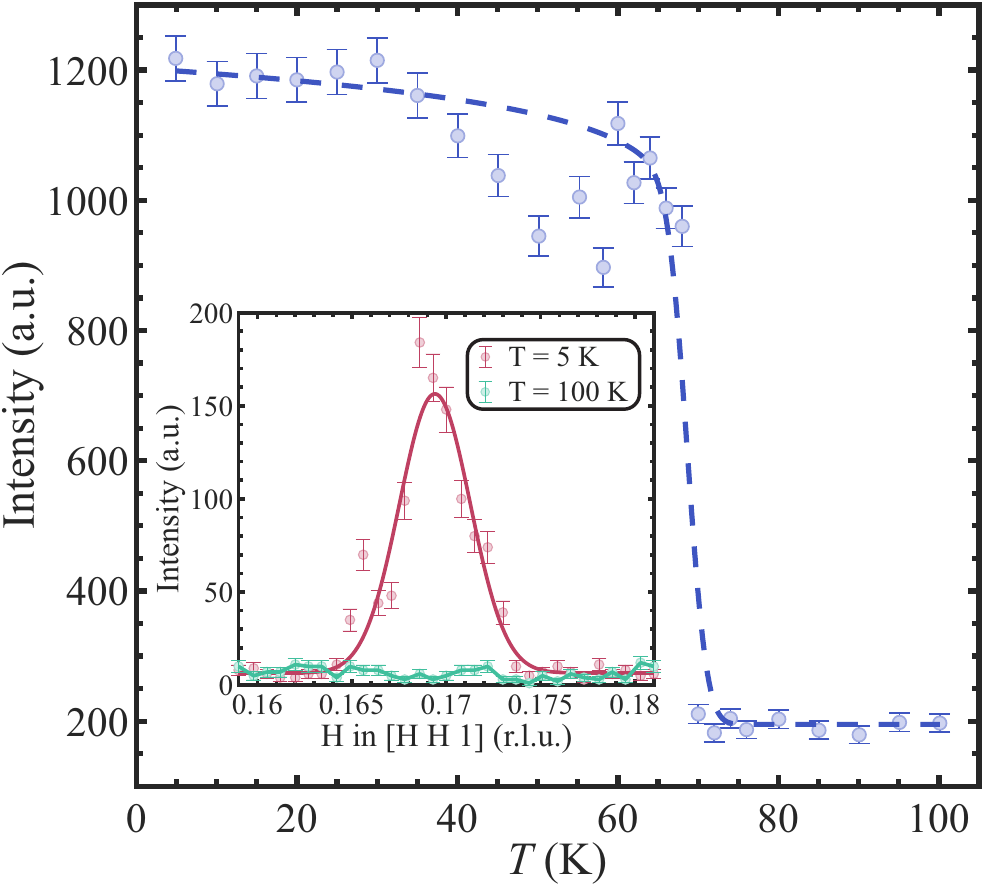}
\caption{Integrated intensity as a function of temperature $T$ of the $(1/6,1/6,1)$ magnetic peak showing a sharp transition at $T=70$ K, consistent with specific-heat measurements in Ref.~\cite{Sangeetha2021}, which reveal a first-order transition at $T_{\rm N}=70$ K. This indicates that the first-order transition in the heat capacity is associated with the magnetic transition. The dashed line is a guide to the eye. The data near $T_{\rm N}=70$ K also indicate a first-order magnetic transition. The inset shows the $(1/6,1/6,1)$ peak at $T = 5$ and 100 K. The weak minimum below $T_{\rm N}$ at $\approx50$ K does not appear in the specific-heat measurements and is currently not understood.}
\label{Fig:OP}
\end{figure}

The temperature dependence of the integrated intensity of the ($\eta$,$\eta$,1) reflections in Fig.~\ref{Fig:OP} shows a very sharp transition at $T=70(1)$ K that coincides with a previous report indicating a strong first-order magnetic phase transition at this temperature \cite{Sangeetha2021}.
The fact that the peak intensities of the $(\pm \eta,\pm \eta,L)$ reflections fall off for larger $L$, as expected from the magnetic form factor of Mn$^{2+}$, is further evidence that these newly observed Bragg peaks are magnetic in origin.
Below, we propose various related magnetic structures that are consistent with the experimental observations assuming the magnetic propagation vector is  $\boldsymbol{\tau}$ = $(\eta,\eta,0)$ r.l.u. with $\eta = 1/6$.

 \begin{figure*}
\centering
\includegraphics[width=\linewidth]{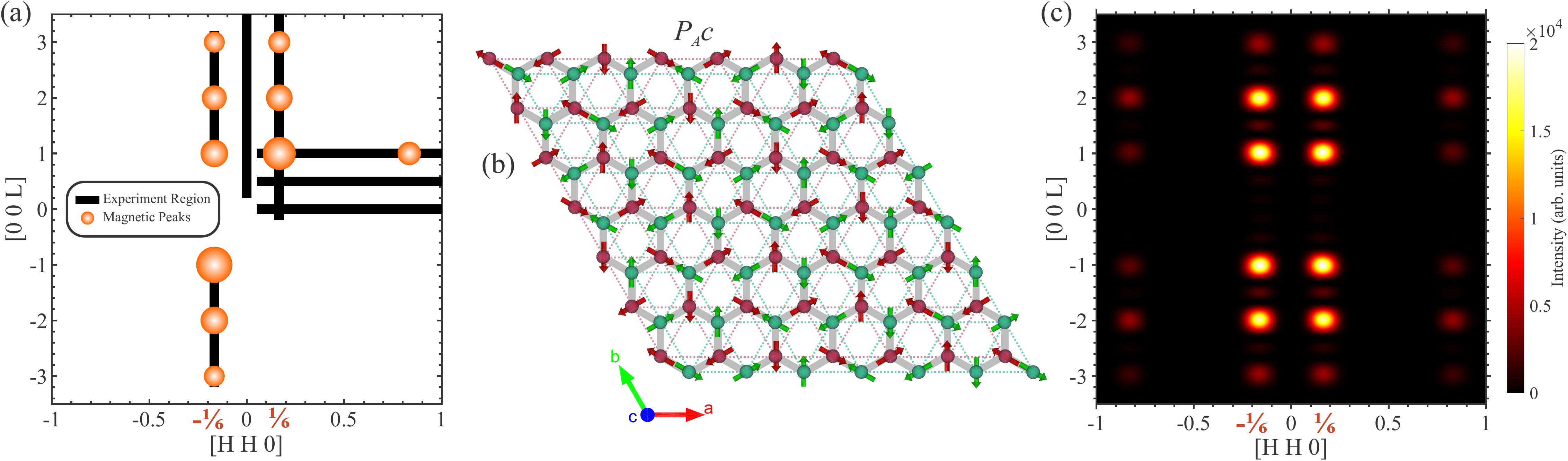}
\caption{(a) A compilation of the magnetic reflections observed in the $(H,H,L)$ planes, where the sizes of the spheres roughly reflect observed intensities. The regions in which the neutron-diffraction experiment was performed are shown by solid black lines. (b)  $P_Ac$ magnetic model structures \cite{bilbao}. The model structures are constructed by creating a $6\times6$ in-plane unit cell consisting of the corrugated honeycomb structure. The sites in green correspond to one trigonal layer (magnetic sublattice) and those in red to the other sublattice. More details on the construction of the magnetic structure are provided in the SM~\cite{SM-CaMn2P2}. (c) Color map of the calculated  structure factor  based on the magnetic structure shown in (b), which is consistent with the experimental results shown in (a).}
\label{Fig:ModelCalc}
\end{figure*} 
\subsection{Analysis of experimental results}
\begin{figure}
\centering
\includegraphics[width=\linewidth]{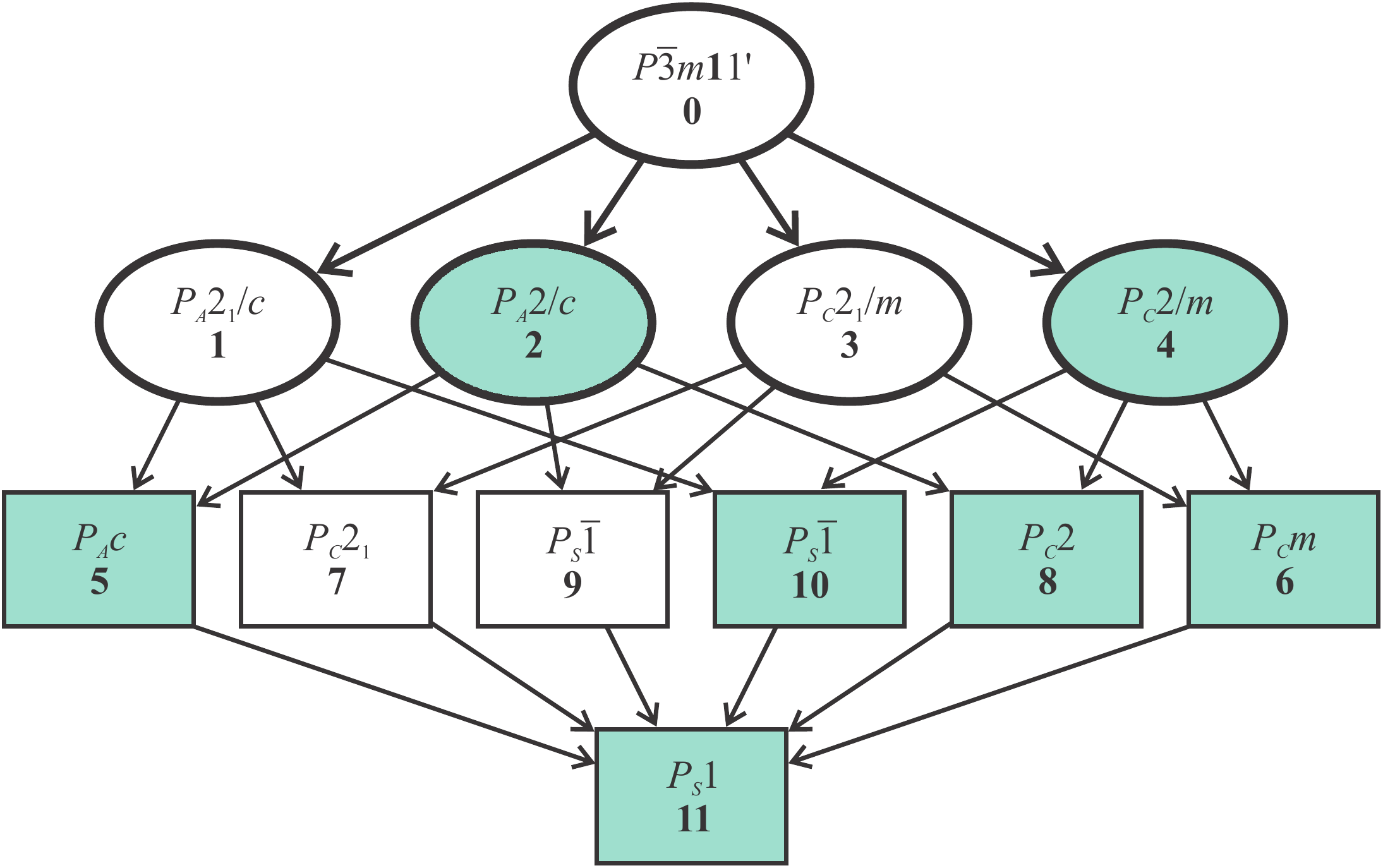}
\caption{Allowed magnetic space groups under the crystallographic space group $P\bar{3}m11^\prime$. Magnetic space groups shaded in green  are the ones that are consistent with our experimentally observed diffraction patterns \cite{bilbao}.}
\label{Fig:msgGroups}
\end{figure}

The observed $(\eta,\eta,0)$ propagation vector indicates that the magnetic structure consists of a $6\times6$ nuclear basal unit cell. Figure \ref{Fig:ModelCalc}(a) is a compilation of the magnetic reflections observed in the $(H,H,L)$ plane, where the sizes of the circles (i.e., peaks) approximate the observed intensities. A systematic analysis reveals that there are seven magnetic space groups (MSGs) that are consistent with the observed magnetic-diffraction patterns. These are $P_A2/c$, $P_C2/m$, $P_Ac$, $P_S\bar{1}$, $P_C2$, $P_Cm$, and $P_S1$ (See Fig. \ref{Fig:msgGroups}). The first two have higher symmetry and $P_Ac$, $P_C2$ are descendants of $P_A2/c$, while $P_S\bar{1}$, $P_C2$, and $P_Cm$ are descendants of $P_C2/m$, and the group $P_S1$ has the lowest symmetry (see Fig.~\ref{Fig:diffraction2} and the Supplemental Material (SM)~\cite{SM-CaMn2P2} for details). 

We now describe an intuitive approach to the magnetic model structure (corresponding to  $P_Ac$), which is constructed by creating a $6\times6$ in-plane nuclear unit cell that spans the corrugated honeycomb structure, i.e., the bilayer magnetic structure stacked along the $c$-axis [Fig.~\ref{Fig:ModelCalc}(b)]. Throughout, the red sites correspond to one trigonal magnetic sublattice and the green sites to the other magnetic sublattice. A magnetic model is constructed by assigning a moment along a high symmetry direction at an origin, for instance, at the lower-left corner, and then successively rotating the spin on the nearest neighbors on the same sublattice clockwise by 60$^{\circ}$. The other sublattice is constructed similarly and stacked with anti-parallel spins with respect to the first sublattice. See more details on the construction of the magnetic structure in the SM~\cite{SM-CaMn2P2}. Note that along the $[1,0,0]$ and $[0,1,0]$ directions, the magnetic structure of each sublattice is a cycloid with a 60$^{\circ}$ turn angle. Thus, for each sublattice, the overall structure is a cycloid with propagation vector $(\eta, \eta, 0)$, with $\eta=1/6$. Inspection of Figure~\ref{Fig:ModelCalc}b
shows that each hexagon consists of two NN antiparallel pairs and one antiparallel NNN pair, such that the net magnetic moment in each hexagon is zero. Also, note that in this model all NN spins along the long diagonal are antiparallel.  

To model the intensities of the magnetic peaks, $I$, we use the following equation:
\begin{equation}
    I =C|f(Q)|^2\left|{\sum_{j=1}^{k}}{\rm e^{i{\bf Q\cdot r_j}}{\bf {\hat Q}}\times({\bf {\hat m_j}}\times{\bf \hat{Q}}})\right|^2,
\label{sfequation}
\end{equation}
where $C$ is a scale factor, ${\bf Q}$ is the scattering vector, ${\bf r}_j$ and ${\bf \hat{m}}_j$ are the position of Mn moment and the unit vector of the magnetic moment, respectively.  $f(Q)$ is the magnetic form factor of Mn$^{2+}$. Using Eq.~(\ref{sfequation}), the calculated magnetic intensities shown in Fig.~\ref{Fig:ModelCalc}(c) are in good agreement with the experimental results shown in Fig.~\ref{Fig:ModelCalc}(a). 

The intensity calculations [Eq.~(\ref{sfequation})] allow us to estimate the average ordered magnetic moment, $\langle gS \rangle$, where $g=2$ is the spectroscopic-splitting factor, $S$ is the spin quantum number, and ${\rm \mu_B}$ is the Bohr magneton. By comparing nuclear-peak intensities and their structure factors to the observed magnetic-peak intensities, we estimate $\langle gS \rangle \mu_{\text{B}} = 4.2(5) \mu_\text{B}$, typical for Mn$^{2+}$ moments.

\section{Theoretical discussion} 
\begin{figure*}[t!]
    \centering
    \includegraphics[width=\linewidth]{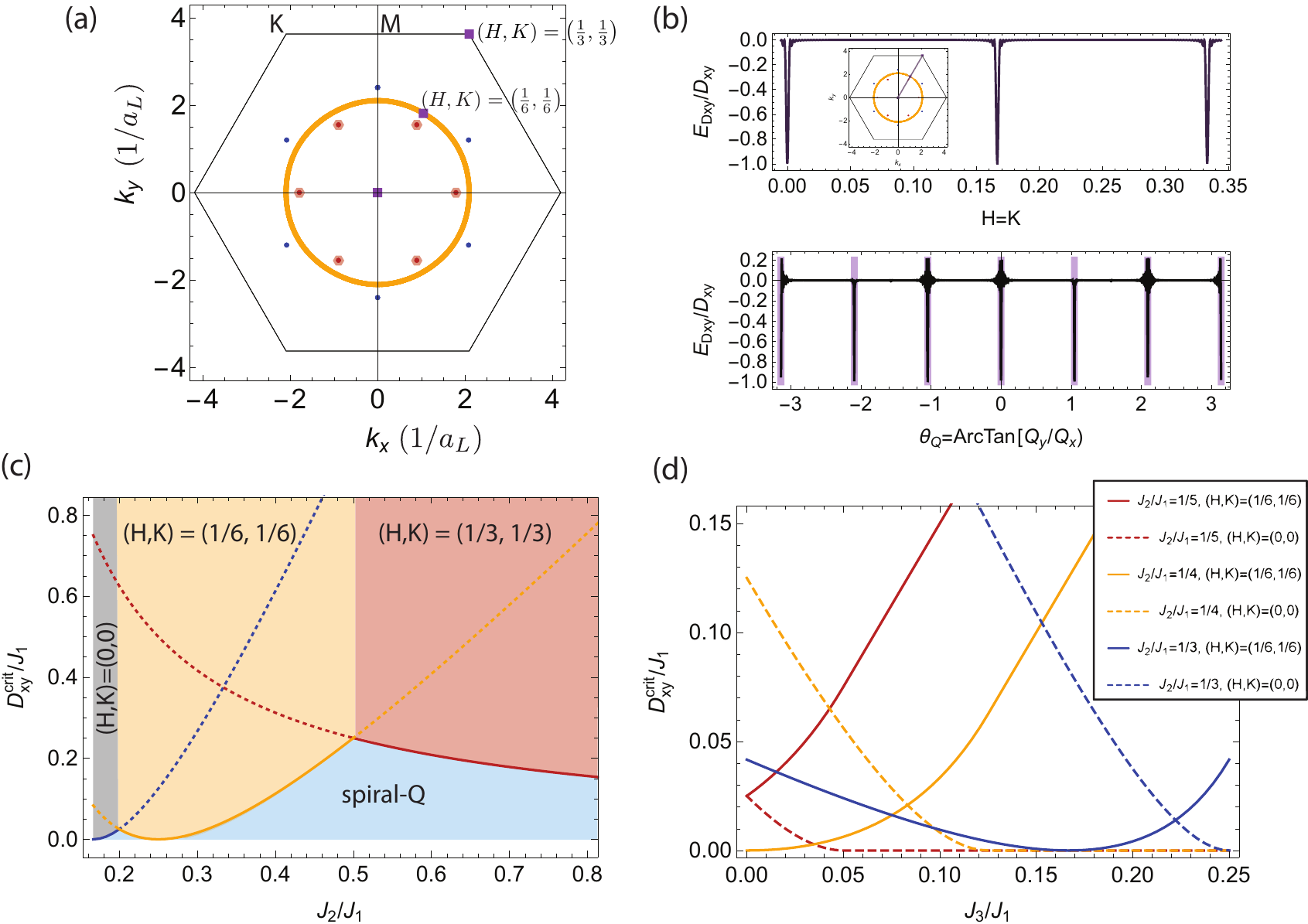}
    \caption{(a) Manifold of wavevectors $(k_x, k_y)$ of spiral magnetic ground-states in the $J_1$-$J_2$-$J_3$ Heisenberg model for $J_2/J_1 = 0.25$. Different colors correspond to different values of $J_3$: $J_3 = 0$ (yellow), AFM $J_3/J_1 = 0.05$ (red hexagons) and FM $J_3/J_1 = -0.05$ (blue dots). 
    Sixfold anisotropy $D_{xy}$ favors wavevectors shown as purple squares (as well as symmetry related ones). 
    (b) Upper panel shows sixfold anisotropy energy $E_{D_{xy}}$ of single-Q magnetic spirals with $H=K$ along a path from the origin to $K$ (shown in inset). Anisotropy favors spirals with $H=K=0$ (N\'eel order), $H=K=\frac16$, and $H=K=\frac13$. Lower panel shows $E_{D_{xy}}$ of single-Q magnetic spirals along yellow circle in (a) as a function of polar angle. Anisotropy favors $H=K=\frac16$ and symmetry-related wavevectors obtained by $60^\circ$ rotations.
    (c) Magnetic ground-state phase diagram of $J_1$-$J_2$-$D_{xy}$ Heisenberg model as a function of $J_2$ and $D_{xy}$ for fixed $J_1 = 1$ and $J_3 = 0$. The solid and dashed lines denote critical $D_{xy}^{\text{crit}}$ that favor commensurate states with $(H,K)=(0,0)$ (blue), $(H,K) = (\frac16, \frac16)$ (yellow) and $(H,K) = (1/3, 1/3)$ (red) over spirals with other wavevectors. The experimentally observed $(H,K) = (\frac16, \frac16)$ spiral phase extends from $0.2 \leq J_2/J_1 \leq 0.5$ and the critical value of $D_{xy}$ exhibits a minimum of zero at $J_2/J_1 =0.25$. 
    (d) Critical value of the sixfold anisotropy $D_{xy}^{\text{crit}}$ as a function of $J_3$ required to favor a commensurate spirals with $(H,K) = (\frac16, \frac16)$ (solid) or $(H,K) = (0,0)$ (dashed). Nonzero $J_3$ moves the ground state wavevector $Q$ closer to the origin, resulting in monotonously decreasing dashed lines. $D_{xy}^{\text{crit}}$ for $(H,K)=(\frac16, \frac16)$ (solid) decreases if $J_3$ moves $Q$ closer to $(\frac16, \frac16)$ (see $J_2/J_1 = 1/3$), and increases otherwise (see $J_2/J_1 = 0, 1/4$). }
    \label{fig:theory_fig}
\end{figure*}

\subsection{Modeling in terms of a Heisenberg Hamiltonian}
\label{subsec:J1J2J3-model}
We interpret the experimental results in the framework of a two-dimensional $J_1$-$J_2$-$J_3$ Heisenberg model including local anisotropy terms on the honeycomb lattice. We find that this model adequately describes the moments on the puckered-honeycomb  Mn$^{2+}$ ions in a single layer of CaMn$_2$P$_2$. Since moments in different layers order ferromagnetically in the three-dimensional crystal, we focus on a single layer in the following. Our model includes NN interactions $J_1$, NNN $J_2$, and third neighbor interactions $J_3$. We also include single-ion anisotropies $D_z$ and $D_{xy}$ that force the moments to lie within the lattice $xy$ plane ($D_z > 0$) and introduce a sixfold in-plane anisotropy ($D_{xy}$), in agreement with the crystalline (point group $\bar{3}m$ or $D_{3d}$) and time-reversal symmetries. Since the orbital moment of Mn$^{2+}$ vanishes according to Hund's rules, the sixfold anisotropy $D_{xy}$ in CaMn$_2$P$_2$ is expected to be small. We model the spins classically, which is well justified given our experimental observation that $\langle gS \rangle \approx 4.3$. 
The Hamiltonian reads
\begin{align}
H &= J_1 \sum_{\langle n,m \rangle_1} \bfss_n \cdot \bfss_m + J_2 \sum_{\langle n,m \rangle_2} \bfss_n \cdot \bfss_m \nonumber \\
& \quad  + J_3 \sum_{\langle n,m \rangle_3} \bfss_n \cdot \bfss_m  + D_z \sum_n (S_n^z)^2 \nonumber \\
& \quad + \frac{D_{xy}}{2} \sum_n \Bigl[ (S_n^x + i S_n^y)^6 + \text{c.c.}  \Bigr] \,,
\label{eq:Heisenberg_Hamiltonian}
\end{align}
where $\bfss_i$ are vectors normalized to $|\bfss_i| = S$, and $n,m$ denote lattice sites of the honeycomb lattice. The summation over $\langle n,m \rangle_\nu$ runs over each $\nu$-th-neighbor bond once. The honeycomb lattice is generated by the triangular Bravais lattice vectors $\bfrr_i = i_1 \bfa_1 + i_2 \bfa_2$ with $i_1, i_2 \in \mathbb{Z}$, $\bfa_1 = (1,0)$ and $\bfa_2 = (-\frac12, \frac{\sqrt{3}}{2})$. Here, we set the Bravais lattice constant $a_L = 1$. The basis sites are $\delta_A = (0,0)$ and $\delta_B = (0, 1/\sqrt{3})$ such that the composite index in Eq.~\eqref{eq:Heisenberg_Hamiltonian} reads $n = (i,\alpha)$ with $\alpha = A, B$.  The reciprocal-lattice vectors are given by $\bfgg_1 = (2 \pi, \frac{2 \pi}{\sqrt{3}})$ and $\bfgg_2 = (0, \frac{4 \pi}{\sqrt{3}})$, and the first Brillouin zone is depicted in Fig.~\ref{fig:theory_fig}(a). To connect to our experimental notation, we write a vector in momentum space as $\bfk = H \bfgg_1 + K \bfgg_2$ such that the $K$-point is located at $(H,K) = (\frac13, \frac13)$ (corners of the BZ) and one of the $M$-points is located at $(H,K) = (0, \frac12)$ (at the center of the BZ edges).

Next, we analyze the classical ground states of Eq.~\eqref{eq:Heisenberg_Hamiltonian} assuming coplanar magnetic order. The ground state phase diagram of the $J_1$-$J_2$-$J_3$ Heisenberg model was derived in Refs.~\cite{Rastelli1979,Katsura1986,Fouet2001}. A coplanar ground state is in agreement with our experimental data and findings in the literature for the $J_1$-$J_2$-$J_3$ model~\cite{Rastelli1979,Katsura1986,Fouet2001}. It can always be favored by a sufficiently-large single-ion anisotropy $D_z$. In the following we assume $D_z > 0$, corresponding to easy-plane anisotropy, forcing the spins to lie in the $ab$ plane. Following Ref.~\cite{Mulder2010}, we parameterize the coplanar spin configuration on the two sublattices as
\begin{subequations}
    \begin{align}
        \bfss_A(\bfrr_i) &= S \bigl( \sin(\bfqq \cdot \bfrr_i), \cos(\bfqq \cdot \bfrr_i)  \bigr) \label{eq:spin_parametrization_A} \\
        \bfss_B(\bfrr_i) &= -S \bigl( \sin(\bfqq \cdot \bfrr_i + \phi), \cos(\bfqq \cdot \bfrr_i + \phi) \bigr) \label{eq:spin_parametrization_B}\,.
    \end{align}
    \label{eq:spin_parametrization}
\end{subequations}
Here, $\phi + \pi$ describes the phase difference between the spins on the $A$ and $B$ sublattices in the same unit cell $\bfrr_i$. Note that Eq.~\eqref{eq:spin_parametrization_B} contains an explicit minus sign such that $\phi = 0$ corresponds to an antiferromagnetic arrangement of $A$ and $B$ spins in the same unit cell. Using this spin parametrization, the classical energy per spin ($N$ = number of spins) reads
\begin{align}
\frac{E}{N S^2} &= -\frac{J_1}{2} \Bigl[ \cos(Q_b - \phi) + \cos(Q_a + Q_b - \phi) -\cos(\phi)\Bigr] \nonumber \\
& + J_2 \Bigl[ \cos(Q_a) + \cos(Q_b) + \cos(Q_a + Q_b) \Bigr] \nonumber \\
& -\frac{J_3}{2} \Bigl[ \cos(Q_a + 2 Q_b - \phi) + \cos(Q_a) \cos(\phi) \Bigr]\,.
\end{align}
Here, $Q_a = \bfqq \cdot \bfa_1$ and $Q_b = \bfqq \cdot \bfa_2$ such that $\bfqq = H \bfgg_1 + K \bfgg_2 = \frac{Q_a}{2 \pi} \bfgg_1 + \frac{Q_b}{2 \pi} \bfgg_2$. We can analytically find the classical ground state energy from the conditions
\begin{equation}
\frac{\partial E}{\partial Q_a} = \frac{\partial E}{\partial Q_b} = \frac{\partial E}{\partial \phi} = 0 \,.
\end{equation}
Let us first discuss the case of $J_3 = D_{xy} = 0$. Then, the ground state exhibits a continuous degeneracy of spiral states with wavevectors $\bfqq = (Q_a, Q_b)$ that fulfill~\cite{Mulder2010}
\begin{equation}
    \cos(Q_a) + \cos(Q_b) + \cos(Q_a + Q_b) = \frac12 \Bigl( \frac{J_1^2}{4 J_2^2} - 3 \Bigr) \,.
\end{equation}
The phase difference $\phi$ is determined by
\begin{equation}
    \sin(\phi) = 2 J_2 \Bigl[ \sin(Q_b) + \sin(Q_a + Q_b) \Bigr] \,.
\end{equation}
For $\frac16 < J_2/J_1 < \frac12$, the manifold of degenerate wavevectors forms a circle around the $\Gamma$ point, as shown in Fig.~\ref{fig:theory_fig}(a) for $J_2/J_1 = 0.25$. The radius of the circle increases continuously with increasing $J_2$. For $J_2/J_1 > 0.5$, the degenerate states are located around the $K$ and $K'$ points, which they approach in the large $J_2$ limit~\cite{Mulder2010}. We refer to the SM for a detailed derivation of these results. In CaMn$_2$P$_2$ we find the propagation vector $(H,K) = (\frac16, \frac16)$, which lies along the $\Gamma$-$K$ direction and corresponds to one of the degenerate states for $J_2/J_1 = 0.25$. This regime of large frustration is thus relevant for CaMn$_2$P$_2$ and will be our focus in the following. 

Nonzero $J_3$ selects a discrete subset of six wavevectors for the ground state spin configuration. For AFM $J_3 > 0$ these lie along the $\Gamma$-$K$ (and symmetry related) directions in the Brillouin zone [see red hexagons in Fig.~\ref{fig:theory_fig}(a)]. In contrast, for FM $J_3 < 0$ these lie along the $\Gamma$-$M$ direction for $J_2/J_1 < 1/2$ [see blue dots in Fig.~\ref{fig:theory_fig}(a)] and along the $K$-$M$ line for $1/2 < J_2/J_1 < 1$ (not shown). The wavevectors shown in Fig.~\ref{fig:theory_fig}(a) are for AFM $J_3/J_1 = 0.05$ (red hexagons) and for FM $J_3/J_1 = -0.05$ (blue circles). Since AFM $J_3$ favors N\'eel order, which is described by $(H,K) = (0,0)$ and $\phi = 0$, the red wavevectors move towards the $\Gamma$ point with increasing AFM $J_3$. In contrast, with increasing FM $J_3<0$ (i.e. more negative values), they move towards the $M$ point. We note that quantum and thermal fluctuations also select six discrete wavevectors, which correspond to the ones favored by FM $J_3$~\cite{Mulder2010}. We therefore conclude that the experimentally-observed wavevector $(H,K) = (\frac16, \frac16)$ is consistent with AFM $J_3>0$. In contrast, it is not favored by FM $J_3$ and it is also not selected via an order-by-disorder mechanism.  

We now analyze the effect of a local sixfold single-ion anisotropy term whose strength is parametrized by $D_{xy}$ [see Eq.~\eqref{eq:Heisenberg_Hamiltonian}]. As shown in Fig.~\ref{fig:theory_fig}(b), nonzero $D_{xy}$ favors a discrete number of spiral states, which are consistent with an alignment of spins along one of the six high symmetry directions on every site. Moving along the direction $H = K$ in the Brillouin zone, we find that $D_{xy}$ equally favors N\'eel order ($H=K=0)$, the experimentally observed spiral order with $H=K=\frac16$ and a shorter spiral with wavevector $H=K=\frac13$ ($K$-point). These three wavevectors are also highlighted in Fig.~\ref{fig:theory_fig}(a) as purple squares. In addition to these three wavevectors, $D_{xy}$ also favors symmetry-related wavevectors as shown in the lower panel of Fig.~\ref{fig:theory_fig}(b), which are obtained by $60^\circ$ rotations. In Fig.~\ref{fig:theory_fig}(c), we show that a magnetic spiral with the experimentally-observed wavevector $(H,K) = (\frac16, \frac16)$ is stabilized over a wide region of $J_2/J_1$ and $D_{xy}$. Specifically, for $0.2 < J_2/J_1< 0.5$ and $J_3 = 0$, the system will enter a magnetic spiral with $H=K = \frac16$ at a critical value of $D_{xy}^{\text{crit}}$ (yellow region). The critical value $D_{xy}^{\text{crit}}$ is a convex function of $J_2/J_1$ and exhibits a minimum of zero at $J_2/J_1 = 0.25$. For smaller values of $J_2/J_1 < 0.2$, the sixfold anisotropy will drive the system into a N\'eel-ordered phase instead (gray region), while for larger values of $J_2/J_1 > 0.5$, it will transition into a magnetic spiral with $H=K = \frac13$ (red region). For nonzero AFM $J_3$ the N\'eel phase extends until larger values of $J_2/J_1$, which sets a limit to the size of $J_3/J_1$ in CaMn$_2$P$_2$. 

To study the dependence on $J_3/J_1$ more systematically, we plot in Fig.~\ref{fig:theory_fig}(d) the evolution of $D_{xy}^{\text{crit}}$ as a function of AFM $J_3/J_1$ for several fixed values of $J_2/J_1$. We focus on the region of $J_2/J_1 < 0.5$, where the N\'eel-ordered phase competes with the $H=K=\frac16$ phase. First, we find that the behavior of $D_{xy}^{\text{crit,1/6}}$ (solid lines) depends on the value of $J_2/J_1$. Since increasing $J_3$ moves the minimum-energy spiral wavevector towards the $\Gamma$ point, $J_3$ reduces $D_{xy}^{\text{crit,1/6}}$ for $J_2/J_1 > 1/4$, but increases it for $J_2/J_1 < 1/4$. Second, since $J_3$ favors the N\'eel ordered state over the spiral, we observe that increasing $J_3$ generally reduces the critical value $D_{xy}^{\text{crit, N\'eel}}$ needed to stabilize the N\'eel phase (dashed lines). The dashed lines are thus monotonously decreasing as a function of $J_3$. For a given value of $J_2/J_1$, we thus find that $D_{xy}^{\text{crit,N\'eel}} < D_{xy}^{\text{crit, 1/6}}$ for sufficiently large $J_3$ such that the sixfold anisotropy drives the system into the N\'eel phase. The position of the crossing point between solid and dashed lines in Fig.~\ref{fig:theory_fig}(d) increases with increasing $J_2/J_1$, which is a result of the minimum-energy wavevector lying closer to $H=K=\frac16$ than to the origin [see Fig.~\ref{fig:theory_fig}(a)]. 

We conclude from this analysis that when $0.2 < J_2/J_1 < 0.5$, the presence of a sixfold anisotropy $D_{xy}$ is sufficient to stabilize the $H=K=\frac16$ spiral order even without a third-neighbor interaction term $J_3$. The required value of $D_{xy}$ to drive the system from an incommensurate spiral into the commensurate $H=K=\frac16$ spiral phase vanishes at $J_2/J_1 = 0.25$ and remains small in the vicinity of this point. Regarding the effect of nonzero $J_3$, we find that AFM $J_3$ selects a wavevector along the observed $H=K$ direction while FM $J_3$ selects different wavevectors that are at $30^\circ$-rotated directions in the Brillouin zone. An AFM third-neighbor interaction is thus more consistent with our experimental findings than a FM one. Since AFM $J_3$ also favors the N\'eel state, the minimum-energy spiral wavevector $\bfqq$ moves towards the origin with increasing $J_3$. For large $J_2/J_1 > 0.5$, where the wavevector lies between the $(\frac16, \frac16)$ and the $K$ point, this moves $\bfqq$ closer to $(\frac16, \frac16)$ and thus reduces the value of $D_{xy}$ necessary to enter the commensurate $H=K=\frac16$ spiral phase [see blue line in Fig.~\ref{fig:theory_fig}(d)]. For smaller values of $J_2/J_1$, a larger value of $J_3$ drives the system into the N\'eel phase and can thus be excluded for CaMn$_2$P$_2$. To summarize, the most likely parameter range describing CaMn$_2$P$_2$ is $J_2/J_1 \approx 0.25 - 0.4$, $J_3/J_1 \lesssim 0.1$ and $D_{xy} > D_{xy}^{\text{crit}} \approx 0 - 0.1 J_1$.

\begin{figure}[t]
    \centering
    \includegraphics[width=.9\linewidth]{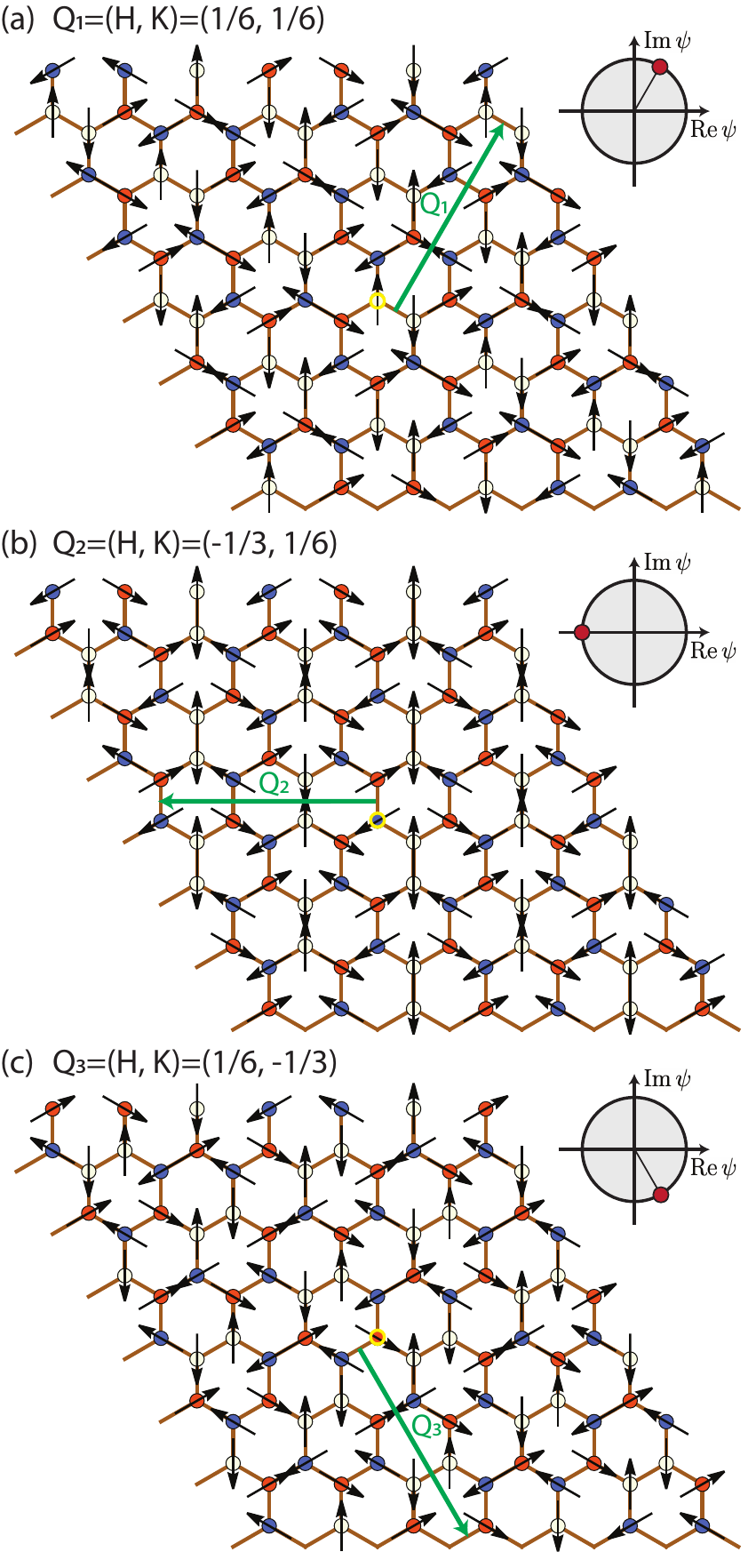}
    \caption{The panels (a,b,c) show the three degenerate ground states $(H,K) =(\frac16, \frac16), (-\frac13, \frac16), (\frac16, -\frac13)$ of the $J_1$-$J_2$-$D_{xy}$ Heisenberg model for $D_{xy} > 0$. In each panel, the bond order parameter $\psi(\bfrr)$ is invariant under translations, but different panels describe different bond orders: the antiparallel nearest-neighbor spin pair occurs along three different bonds in the three panels (a-c). The complex argument of $\psi$ is given by the polar angle of the corresponding wavevector $\bfqq_i$ in the Brillouin zone (see red hexagons in Fig.~\ref{fig:theory_fig}).}
    \label{fig:theory_Potts}
\end{figure}

\subsection{Emergent Potts-nematic order and first-order phase transition}
\label{subsec:emergent_Potts}
The frustration-induced spiral magnetic order that we observe in CaMn$_2$P$_2$ leads to the emergence of a Potts-nematic order parameter. This composite order parameter is bilinear in the spins and involves their scalar product on nearest-neighbor sites: 
\begin{align}
        \psi(\bfrr) &= \bfss_A(\bfrr) \cdot \bfss_B(\bfrr) + e^{-\frac{2 \pi i}{3}} \bfss_A(\bfrr) \cdot \bfss_B(\bfrr - \bfa_2) \nonumber \\ & \qquad + e^{-\frac{4 \pi i}{3}} \bfss_A(\bfrr) \cdot \bfss_B(\bfrr - \bfa_1 - \bfa_2) \,.
    \label{eq:psi_Potts}
\end{align}
This complex bond order parameter is finite and translationally invariant in any of the three spiral magnetic states with $(H,K) = \{\bfqq_1, \bfqq_2, \bfqq_3\}=(\frac16, \frac16), (-\frac13, \frac16), (\frac16, -\frac13)$. In Fig.~\ref{fig:theory_Potts}, we show the three degenerate ground states of the $J_1$-$J_2$-$J_3$-$D_{xy}$ Heisenberg model in the regime where $D_{xy} > D_{xy,\text{crit}}$ and $0.2 < J_2/J_1 < 0.5$ [see Fig.~\ref{fig:theory_fig}(c)]. When placed on the Mn ions in CaMn$_2$P$_2$, this magnetic structure corresponds to magnetic space group (MSG) $P_S1$, which is one of the MSGs that are consistent with experiment (see Fig.~\ref{Fig:msgGroups}). The related magnetic structure for $D_{xy} < 0$, for which the spin at the origin (yellow circle) is rotated by $\frac{\pi}{2}$, lies in the MSG $P_Cm$ that is also consistent with the experimental data. 
The three panels in Fig.~\ref{fig:theory_Potts} depict the symmetry-related states with propagation vectors $\bfqq_i$ and the insets show the value of the spatially homogeneous complex Potts-nematic order parameter, whose argument follows the direction of the ordering wavevector. It is a generalization of the Ising nematic bond order parameter known to underlie the tetragonal to orthorhombic transition via magneto-elastic couplings that is observed in tetragonal iron-based arsenides such as CaFe$_2$As$_2$~\cite{Ni2008,Fernandes2012}.

Under a threefold rotation around an $A$ site, the Potts-nematic order parameter transforms as $\psi \xrightarrow{C_3} \exp\bigl(\frac{2 \pi i}{3}\bigr) \psi$. Under a mirror operation $m_{yz}$ that sends $x \rightarrow -x$, it transforms as $\psi \xrightarrow{m_{yz}} \psi^*$. Its finite temperature behavior is thus described by the Landau-Ginzburg free energy functional of a three-state Potts model~\cite{Mulder2010}. In three dimensions this analysis predicts a first-order phase transition into a state with long-range Potts order due to a symmetry-allowed third-order term. This is also in agreement with Monte-Carlo simulations~\cite{Janke1997}. The emergence of long-range Potts-nematic order can therefore naturally account for the experimentally-observed first-order magnetic phase transition in CaMn$_2$P$_2$. Note that different honeycomb layers are ordered ferromagnetically along the $c$ direction in CaMn$_2$P$_2$, corresponding to an ordering wavevector with integer $L$, where $\bfqq = (H,K,L)$. Since $\psi$ is a composite magnetic order parameter, it is strongly intertwined with magnetism, and the discontinuous development of long-range Potts order at the first-order transition can thus uplift the magnetic transition to occur as a joint first-order transition. The system then simultaneously develops long-range Potts-nematic and magnetic order. Such a behavior is known to occur, for example, in the triangular lattice antiferromagnet Fe$_{1/3}$NbSe$_2$~\cite{Little2020}. This can also explain why the related compounds CaMn$_2$$Pn_2$ with $Pn$ = Sb, Bi that exhibit N\'eel order, for which such a 3-state Potts-nematic order is absent, develop magnetic order via a continuous phase transition. 

Since long-range Potts-nematic order breaks the threefold rotation symmetry of the lattice, we predict the emergence of three lattice distortion domains due to a finite magneto-elastic coupling. The domains are characterized by different values of the Potts-nematic order parameter $\psi$, as shown in Fig.~\ref{fig:theory_Potts}. However, the coupling between magnetic and lattice degrees of freedom is expected to be small in this system, since the orbital moment of the magnetic ions Mn$^{2+}$ vanishes according to the Hund's rules and spin-orbit coupling is therefore small. This could be the reason why lattice distortion and crystal symmetry lowering could not be detected in previous x-ray diffraction studies~\cite{Sangeetha2021}. An alternative explanation is the emergence of a complex multi-$Q$ magnetic order that preserves all lattice symmetries. It is worth noting, however, that Raman scattering studies have reported the appearance of additional peaks when going from the paramagnetic phase at room temperature to the magnetic phase at $T < T_N$~\cite{Li2020}. Further investigations of the effects of magnetic ordering on the lattice and its excitations are needed to address these open questions. We emphasize that magnetic spiral-$Q$ order with a single finite-momentum wavevector breaks threefold rotation symmetry via selection of one of the three symmetry-equivalent propagation vectors $\bfqq_i$. In the single-$Q$ spiral state we thus expect the appearance of three magnetic domains characterized by different magnetic propagation vectors in the magnetically ordered state.

\section{Conclusions}
\label{sec:conclusions}
Using neutron-diffraction measurements, we find that CaMn$_2$P$_2$ undergoes a first-order antiferromagnetic transition at $T_{\rm N}=70$ K into a state with a $6\times 6$ times enlarged magnetic unit cell. The average ordered magnetic moment is $\langle gS \rangle \mu_{\text{B}} = 4.2(5)$ ${\rm \mu}_{\rm B}$. The integrated intensity of the major $(H,K,L) = (\frac16, \frac16,1)$ magnetic peak versus temperature shows an abrupt decrease in intensity at $T_{\rm N}$ that is a characteristic of a first-order phase transition. Focusing on the experimentally discovered ground-state, we interpret these results using a frustrated $J_1$-$J_2$-$J_3$ Heisenberg model with easy-plane anisotropy $D_z$ and a sixfold in-plane anisotropy $D_{xy}$, and show that this propagation wavevector signals the presence of a substantial degree of frustration. We relate the appearance of the first-order magnetic transition to a composite three-state Potts-nematic bond order parameter that simultaneously develops long-range order and drives the magnetic transition to become first-order. Based on our analysis we predict the emergence of three symmetry related magnetic and lattice distortion domains that deserve further studies. 


\acknowledgments
This research was supported by the U.S. Department of Energy, Office of Basic Energy Sciences, Division of Materials Sciences and Engineering. Ames Laboratory is operated for the U.S. Department of Energy by Iowa State University under Contract No.~DE-AC02-07CH11358.

\bibliography{bib_zotero}

\begin{thebibliography}{44}%
\makeatletter
\providecommand \@ifxundefined [1]{%
 \@ifx{#1\undefined}
}%
\providecommand \@ifnum [1]{%
 \ifnum #1\expandafter \@firstoftwo
 \else \expandafter \@secondoftwo
 \fi
}%
\providecommand \@ifx [1]{%
 \ifx #1\expandafter \@firstoftwo
 \else \expandafter \@secondoftwo
 \fi
}%
\providecommand \natexlab [1]{#1}%
\providecommand \enquote  [1]{``#1''}%
\providecommand \bibnamefont  [1]{#1}%
\providecommand \bibfnamefont [1]{#1}%
\providecommand \citenamefont [1]{#1}%
\providecommand \href@noop [0]{\@secondoftwo}%
\providecommand \href [0]{\begingroup \@sanitize@url \@href}%
\providecommand \@href[1]{\@@startlink{#1}\@@href}%
\providecommand \@@href[1]{\endgroup#1\@@endlink}%
\providecommand \@sanitize@url [0]{\catcode `\\12\catcode `\$12\catcode
  `\&12\catcode `\#12\catcode `\^12\catcode `\_12\catcode `\%12\relax}%
\providecommand \@@startlink[1]{}%
\providecommand \@@endlink[0]{}%
\providecommand \url  [0]{\begingroup\@sanitize@url \@url }%
\providecommand \@url [1]{\endgroup\@href {#1}{\urlprefix }}%
\providecommand \urlprefix  [0]{URL }%
\providecommand \Eprint [0]{\href }%
\providecommand \doibase [0]{https://doi.org/}%
\providecommand \selectlanguage [0]{\@gobble}%
\providecommand \bibinfo  [0]{\@secondoftwo}%
\providecommand \bibfield  [0]{\@secondoftwo}%
\providecommand \translation [1]{[#1]}%
\providecommand \BibitemOpen [0]{}%
\providecommand \bibitemStop [0]{}%
\providecommand \bibitemNoStop [0]{.\EOS\space}%
\providecommand \EOS [0]{\spacefactor3000\relax}%
\providecommand \BibitemShut  [1]{\csname bibitem#1\endcsname}%
\let\auto@bib@innerbib\@empty
\bibitem [{\citenamefont {Chaloupka}\ \emph {et~al.}(2010)\citenamefont
  {Chaloupka}, \citenamefont {Jackeli},\ and\ \citenamefont
  {Khaliullin}}]{Chaloupka2010}%
  \BibitemOpen
  \bibfield  {author} {\bibinfo {author} {\bibfnamefont {J.}~\bibnamefont
  {Chaloupka}}, \bibinfo {author} {\bibfnamefont {G.}~\bibnamefont {Jackeli}},\
  and\ \bibinfo {author} {\bibfnamefont {G.}~\bibnamefont {Khaliullin}},\
  }\bibfield  {title} {\bibinfo {title} {Kitaev-{{Heisenberg Model}} on a
  {{Honeycomb Lattice}}: {{Possible Exotic Phases}} in {{Iridium Oxides
  A}}{\textsubscript{2}}{{IrO}}{\textsubscript{3}}},\ }\href
  {https://doi.org/10.1103/PhysRevLett.105.027204} {\bibfield  {journal}
  {\bibinfo  {journal} {Phys. Rev. Lett.}\ }\textbf {\bibinfo {volume} {105}},\
  \bibinfo {pages} {027204} (\bibinfo {year} {2010})}\BibitemShut {NoStop}%
\bibitem [{\citenamefont {Trebst}\ and\ \citenamefont
  {Hickey}(2022)}]{Trebst2022}%
  \BibitemOpen
  \bibfield  {author} {\bibinfo {author} {\bibfnamefont {S.}~\bibnamefont
  {Trebst}}\ and\ \bibinfo {author} {\bibfnamefont {C.}~\bibnamefont
  {Hickey}},\ }\bibfield  {title} {\bibinfo {title} {Kitaev materials},\ }\href
  {https://doi.org/10.1016/j.physrep.2021.11.003} {\bibfield  {journal}
  {\bibinfo  {journal} {Physics Reports}\ }\textbf {\bibinfo {volume} {950}},\
  \bibinfo {pages} {1} (\bibinfo {year} {2022})}\BibitemShut {NoStop}%
\bibitem [{\citenamefont {Morey}\ \emph {et~al.}(2019)\citenamefont {Morey},
  \citenamefont {Scheie}, \citenamefont {Sheckelton}, \citenamefont {Brown},\
  and\ \citenamefont {McQueen}}]{Morey2019}%
  \BibitemOpen
  \bibfield  {author} {\bibinfo {author} {\bibfnamefont {J.~R.}\ \bibnamefont
  {Morey}}, \bibinfo {author} {\bibfnamefont {A.}~\bibnamefont {Scheie}},
  \bibinfo {author} {\bibfnamefont {J.~P.}\ \bibnamefont {Sheckelton}},
  \bibinfo {author} {\bibfnamefont {C.~M.}\ \bibnamefont {Brown}},\ and\
  \bibinfo {author} {\bibfnamefont {T.~M.}\ \bibnamefont {McQueen}},\
  }\bibfield  {title} {\bibinfo {title}
  {Ni{\textsubscript{2}}{{Mo}}{\textsubscript{3}}{{O}}{\textsubscript{8}} :
  {{Complex}} antiferromagnetic order on a honeycomb lattice},\ }\href
  {https://doi.org/10.1103/PhysRevMaterials.3.014410} {\bibfield  {journal}
  {\bibinfo  {journal} {Phys. Rev. Materials}\ }\textbf {\bibinfo {volume}
  {3}},\ \bibinfo {pages} {014410} (\bibinfo {year} {2019})}\BibitemShut
  {NoStop}%
\bibitem [{\citenamefont {Yehia}\ \emph {et~al.}(2010)\citenamefont {Yehia},
  \citenamefont {Vavilova}, \citenamefont {M{\"o}ller}, \citenamefont {Taetz},
  \citenamefont {L{\"o}w}, \citenamefont {Klingeler}, \citenamefont {Kataev},\
  and\ \citenamefont {B{\"u}chner}}]{Yehia2010}%
  \BibitemOpen
  \bibfield  {author} {\bibinfo {author} {\bibfnamefont {M.}~\bibnamefont
  {Yehia}}, \bibinfo {author} {\bibfnamefont {E.}~\bibnamefont {Vavilova}},
  \bibinfo {author} {\bibfnamefont {A.}~\bibnamefont {M{\"o}ller}}, \bibinfo
  {author} {\bibfnamefont {T.}~\bibnamefont {Taetz}}, \bibinfo {author}
  {\bibfnamefont {U.}~\bibnamefont {L{\"o}w}}, \bibinfo {author} {\bibfnamefont
  {R.}~\bibnamefont {Klingeler}}, \bibinfo {author} {\bibfnamefont
  {V.}~\bibnamefont {Kataev}},\ and\ \bibinfo {author} {\bibfnamefont
  {B.}~\bibnamefont {B{\"u}chner}},\ }\bibfield  {title} {\bibinfo {title}
  {Finite-size effects and magnetic order in the spin- 1 2 honeycomb-lattice
  compound
  {{InCu}}{\textsubscript{2/3}}{{V}}{\textsubscript{1/3}}{{O}}{\textsubscript{3}}},\
  }\href {https://doi.org/10.1103/PhysRevB.81.060414} {\bibfield  {journal}
  {\bibinfo  {journal} {Phys. Rev. B}\ }\textbf {\bibinfo {volume} {81}},\
  \bibinfo {pages} {060414} (\bibinfo {year} {2010})}\BibitemShut {NoStop}%
\bibitem [{\citenamefont {Iakovleva}\ \emph {et~al.}(2019)\citenamefont
  {Iakovleva}, \citenamefont {Janson}, \citenamefont {Grafe}, \citenamefont
  {Dioguardi}, \citenamefont {Maeter}, \citenamefont {Yeche}, \citenamefont
  {Klauss}, \citenamefont {Pascua}, \citenamefont {Luetkens}, \citenamefont
  {M{\"o}ller}, \citenamefont {B{\"u}chner}, \citenamefont {Kataev},\ and\
  \citenamefont {Vavilova}}]{Iakovleva2019}%
  \BibitemOpen
  \bibfield  {author} {\bibinfo {author} {\bibfnamefont {M.}~\bibnamefont
  {Iakovleva}}, \bibinfo {author} {\bibfnamefont {O.}~\bibnamefont {Janson}},
  \bibinfo {author} {\bibfnamefont {H.-J.}\ \bibnamefont {Grafe}}, \bibinfo
  {author} {\bibfnamefont {A.~P.}\ \bibnamefont {Dioguardi}}, \bibinfo {author}
  {\bibfnamefont {H.}~\bibnamefont {Maeter}}, \bibinfo {author} {\bibfnamefont
  {N.}~\bibnamefont {Yeche}}, \bibinfo {author} {\bibfnamefont {H.-H.}\
  \bibnamefont {Klauss}}, \bibinfo {author} {\bibfnamefont {G.}~\bibnamefont
  {Pascua}}, \bibinfo {author} {\bibfnamefont {H.}~\bibnamefont {Luetkens}},
  \bibinfo {author} {\bibfnamefont {A.}~\bibnamefont {M{\"o}ller}}, \bibinfo
  {author} {\bibfnamefont {B.}~\bibnamefont {B{\"u}chner}}, \bibinfo {author}
  {\bibfnamefont {V.}~\bibnamefont {Kataev}},\ and\ \bibinfo {author}
  {\bibfnamefont {E.}~\bibnamefont {Vavilova}},\ }\bibfield  {title} {\bibinfo
  {title} {Ground state and low-temperature magnetism of the
  quasi-two-dimensional honeycomb compound
  {{InCu}}{\textsubscript{2/3}}{{V}}{\textsubscript{1/3}}{{O}}{\textsubscript{3}}},\
  }\href {https://doi.org/10.1103/PhysRevB.100.144442} {\bibfield  {journal}
  {\bibinfo  {journal} {Phys. Rev. B}\ }\textbf {\bibinfo {volume} {100}},\
  \bibinfo {pages} {144442} (\bibinfo {year} {2019})}\BibitemShut {NoStop}%
\bibitem [{\citenamefont {Smirnova}\ \emph {et~al.}(2009)\citenamefont
  {Smirnova}, \citenamefont {Azuma}, \citenamefont {Kumada}, \citenamefont
  {Kusano}, \citenamefont {Matsuda}, \citenamefont {Shimakawa}, \citenamefont
  {Takei}, \citenamefont {Yonesaki},\ and\ \citenamefont
  {Kinomura}}]{Smirnova2009}%
  \BibitemOpen
  \bibfield  {author} {\bibinfo {author} {\bibfnamefont {O.}~\bibnamefont
  {Smirnova}}, \bibinfo {author} {\bibfnamefont {M.}~\bibnamefont {Azuma}},
  \bibinfo {author} {\bibfnamefont {N.}~\bibnamefont {Kumada}}, \bibinfo
  {author} {\bibfnamefont {Y.}~\bibnamefont {Kusano}}, \bibinfo {author}
  {\bibfnamefont {M.}~\bibnamefont {Matsuda}}, \bibinfo {author} {\bibfnamefont
  {Y.}~\bibnamefont {Shimakawa}}, \bibinfo {author} {\bibfnamefont
  {T.}~\bibnamefont {Takei}}, \bibinfo {author} {\bibfnamefont
  {Y.}~\bibnamefont {Yonesaki}},\ and\ \bibinfo {author} {\bibfnamefont
  {N.}~\bibnamefont {Kinomura}},\ }\bibfield  {title} {\bibinfo {title}
  {Synthesis, {{Crystal Structure}}, and {{Magnetic Properties}} of
  {{Bi}}{\textsubscript{3}}{{Mn}}{\textsubscript{4}}{{O}}{\textsubscript{12}}({{NO}}{\textsubscript{3}})
  {{Oxynitrate Comprising}} {{{\emph{S}}}} = 3/2 {{Honeycomb Lattice}}},\
  }\href {https://doi.org/10.1021/ja901922p} {\bibfield  {journal} {\bibinfo
  {journal} {J. Am. Chem. Soc.}\ }\textbf {\bibinfo {volume} {131}},\ \bibinfo
  {pages} {8313} (\bibinfo {year} {2009})}\BibitemShut {NoStop}%
\bibitem [{\citenamefont {Miyamoto}\ \emph {et~al.}(2019)\citenamefont
  {Miyamoto}, \citenamefont {Iwasaki}, \citenamefont {Uemoto}, \citenamefont
  {Hosokoshi}, \citenamefont {Fujiwara}, \citenamefont {Shimono},\ and\
  \citenamefont {Yamaguchi}}]{Miyamoto2019}%
  \BibitemOpen
  \bibfield  {author} {\bibinfo {author} {\bibfnamefont {S.}~\bibnamefont
  {Miyamoto}}, \bibinfo {author} {\bibfnamefont {Y.}~\bibnamefont {Iwasaki}},
  \bibinfo {author} {\bibfnamefont {N.}~\bibnamefont {Uemoto}}, \bibinfo
  {author} {\bibfnamefont {Y.}~\bibnamefont {Hosokoshi}}, \bibinfo {author}
  {\bibfnamefont {H.}~\bibnamefont {Fujiwara}}, \bibinfo {author}
  {\bibfnamefont {S.}~\bibnamefont {Shimono}},\ and\ \bibinfo {author}
  {\bibfnamefont {H.}~\bibnamefont {Yamaguchi}},\ }\bibfield  {title} {\bibinfo
  {title} {Magnetic properties of honeycomb-based spin models in verdazyl-based
  salts},\ }\href {https://doi.org/10.1103/PhysRevMaterials.3.064410}
  {\bibfield  {journal} {\bibinfo  {journal} {Phys. Rev. Materials}\ }\textbf
  {\bibinfo {volume} {3}},\ \bibinfo {pages} {064410} (\bibinfo {year}
  {2019})}\BibitemShut {NoStop}%
\bibitem [{\citenamefont {Kitaev}(2006)}]{Kitaev2006}%
  \BibitemOpen
  \bibfield  {author} {\bibinfo {author} {\bibfnamefont {A.}~\bibnamefont
  {Kitaev}},\ }\bibfield  {title} {\bibinfo {title} {Anyons in an exactly
  solved model and beyond},\ }\href {https://doi.org/10.1016/j.aop.2005.10.005}
  {\bibfield  {journal} {\bibinfo  {journal} {Ann. Phys.}\ }\textbf {\bibinfo
  {volume} {321}},\ \bibinfo {pages} {2} (\bibinfo {year} {2006})}\BibitemShut
  {NoStop}%
\bibitem [{\citenamefont {Baskaran}\ \emph {et~al.}(2008)\citenamefont
  {Baskaran}, \citenamefont {Sen},\ and\ \citenamefont
  {Shankar}}]{Baskaran2008}%
  \BibitemOpen
  \bibfield  {author} {\bibinfo {author} {\bibfnamefont {G.}~\bibnamefont
  {Baskaran}}, \bibinfo {author} {\bibfnamefont {D.}~\bibnamefont {Sen}},\ and\
  \bibinfo {author} {\bibfnamefont {R.}~\bibnamefont {Shankar}},\ }\bibfield
  {title} {\bibinfo {title} {Spin-{{{\emph{S}}}} {{Kitaev}} model:
  {{Classical}} ground states, order from disorder, and exact correlation
  functions},\ }\href {https://doi.org/10.1103/PhysRevB.78.115116} {\bibfield
  {journal} {\bibinfo  {journal} {Phys. Rev. B}\ }\textbf {\bibinfo {volume}
  {78}},\ \bibinfo {pages} {115116} (\bibinfo {year} {2008})}\BibitemShut
  {NoStop}%
\bibitem [{\citenamefont {Reuther}\ \emph
  {et~al.}(2011{\natexlab{a}})\citenamefont {Reuther}, \citenamefont
  {Thomale},\ and\ \citenamefont {Trebst}}]{Reuther2011}%
  \BibitemOpen
  \bibfield  {author} {\bibinfo {author} {\bibfnamefont {J.}~\bibnamefont
  {Reuther}}, \bibinfo {author} {\bibfnamefont {R.}~\bibnamefont {Thomale}},\
  and\ \bibinfo {author} {\bibfnamefont {S.}~\bibnamefont {Trebst}},\
  }\bibfield  {title} {\bibinfo {title} {Finite-temperature phase diagram of
  the {{Heisenberg-Kitaev}} model},\ }\href
  {https://doi.org/10.1103/PhysRevB.84.100406} {\bibfield  {journal} {\bibinfo
  {journal} {Phys. Rev. B}\ }\textbf {\bibinfo {volume} {84}},\ \bibinfo
  {pages} {100406} (\bibinfo {year} {2011}{\natexlab{a}})}\BibitemShut
  {NoStop}%
\bibitem [{\citenamefont {Price}\ and\ \citenamefont
  {Perkins}(2013)}]{Price2013}%
  \BibitemOpen
  \bibfield  {author} {\bibinfo {author} {\bibfnamefont {C.}~\bibnamefont
  {Price}}\ and\ \bibinfo {author} {\bibfnamefont {N.~B.}\ \bibnamefont
  {Perkins}},\ }\bibfield  {title} {\bibinfo {title} {Finite-temperature phase
  diagram of the classical {{Kitaev-Heisenberg}} model},\ }\href
  {https://doi.org/10.1103/PhysRevB.88.024410} {\bibfield  {journal} {\bibinfo
  {journal} {Phys. Rev. B}\ }\textbf {\bibinfo {volume} {88}},\ \bibinfo
  {pages} {024410} (\bibinfo {year} {2013})}\BibitemShut {NoStop}%
\bibitem [{\citenamefont {Rastelli}\ \emph {et~al.}(1979)\citenamefont
  {Rastelli}, \citenamefont {Tassi},\ and\ \citenamefont
  {Reatto}}]{Rastelli1979}%
  \BibitemOpen
  \bibfield  {author} {\bibinfo {author} {\bibfnamefont {E.}~\bibnamefont
  {Rastelli}}, \bibinfo {author} {\bibfnamefont {A.}~\bibnamefont {Tassi}},\
  and\ \bibinfo {author} {\bibfnamefont {L.}~\bibnamefont {Reatto}},\
  }\bibfield  {title} {\bibinfo {title} {Non-simple magnetic order for simple
  {{Hamiltonians}}},\ }\href {https://doi.org/10.1016/0378-4363(79)90002-0}
  {\bibfield  {journal} {\bibinfo  {journal} {Physica B+C}\ }\textbf {\bibinfo
  {volume} {97}},\ \bibinfo {pages} {1} (\bibinfo {year} {1979})}\BibitemShut
  {NoStop}%
\bibitem [{\citenamefont {Katsura}\ \emph {et~al.}(1986)\citenamefont
  {Katsura}, \citenamefont {Ide},\ and\ \citenamefont {Morita}}]{Katsura1986}%
  \BibitemOpen
  \bibfield  {author} {\bibinfo {author} {\bibfnamefont {S.}~\bibnamefont
  {Katsura}}, \bibinfo {author} {\bibfnamefont {T.}~\bibnamefont {Ide}},\ and\
  \bibinfo {author} {\bibfnamefont {T.}~\bibnamefont {Morita}},\ }\bibfield
  {title} {\bibinfo {title} {The ground states of the classical heisenberg and
  planar models on the triangular and plane hexagonal lattices},\ }\href
  {https://doi.org/10.1007/BF01127717} {\bibfield  {journal} {\bibinfo
  {journal} {J Stat Phys}\ }\textbf {\bibinfo {volume} {42}},\ \bibinfo {pages}
  {381} (\bibinfo {year} {1986})}\BibitemShut {NoStop}%
\bibitem [{\citenamefont {Fouet}\ \emph {et~al.}(2001)\citenamefont {Fouet},
  \citenamefont {Sindzingre},\ and\ \citenamefont {Lhuillier}}]{Fouet2001}%
  \BibitemOpen
  \bibfield  {author} {\bibinfo {author} {\bibfnamefont {J.}~\bibnamefont
  {Fouet}}, \bibinfo {author} {\bibfnamefont {P.}~\bibnamefont {Sindzingre}},\
  and\ \bibinfo {author} {\bibfnamefont {C.}~\bibnamefont {Lhuillier}},\
  }\bibfield  {title} {\bibinfo {title} {An investigation of the quantum
  {{J}}{\textsubscript{1}}-{{J}}{\textsubscript{2}}-{{J}}{\textsubscript{3}}
  model on the honeycomb lattice},\ }\href
  {https://doi.org/10.1007/s100510170273} {\bibfield  {journal} {\bibinfo
  {journal} {Eur. Phys. J. B}\ }\textbf {\bibinfo {volume} {20}},\ \bibinfo
  {pages} {241} (\bibinfo {year} {2001})}\BibitemShut {NoStop}%
\bibitem [{\citenamefont {Mulder}\ \emph {et~al.}(2010)\citenamefont {Mulder},
  \citenamefont {Ganesh}, \citenamefont {Capriotti},\ and\ \citenamefont
  {Paramekanti}}]{Mulder2010}%
  \BibitemOpen
  \bibfield  {author} {\bibinfo {author} {\bibfnamefont {A.}~\bibnamefont
  {Mulder}}, \bibinfo {author} {\bibfnamefont {R.}~\bibnamefont {Ganesh}},
  \bibinfo {author} {\bibfnamefont {L.}~\bibnamefont {Capriotti}},\ and\
  \bibinfo {author} {\bibfnamefont {A.}~\bibnamefont {Paramekanti}},\
  }\bibfield  {title} {\bibinfo {title} {Spiral order by disorder and lattice
  nematic order in a frustrated {{Heisenberg}} antiferromagnet on the honeycomb
  lattice},\ }\href {https://doi.org/10.1103/PhysRevB.81.214419} {\bibfield
  {journal} {\bibinfo  {journal} {Phys. Rev. B}\ }\textbf {\bibinfo {volume}
  {81}},\ \bibinfo {pages} {214419} (\bibinfo {year} {2010})}\BibitemShut
  {NoStop}%
\bibitem [{\citenamefont {Albuquerque}\ \emph {et~al.}(2011)\citenamefont
  {Albuquerque}, \citenamefont {Schwandt}, \citenamefont {Het{\'e}nyi},
  \citenamefont {Capponi}, \citenamefont {Mambrini},\ and\ \citenamefont
  {L{\"a}uchli}}]{Albuquerque2011}%
  \BibitemOpen
  \bibfield  {author} {\bibinfo {author} {\bibfnamefont {A.~F.}\ \bibnamefont
  {Albuquerque}}, \bibinfo {author} {\bibfnamefont {D.}~\bibnamefont
  {Schwandt}}, \bibinfo {author} {\bibfnamefont {B.}~\bibnamefont
  {Het{\'e}nyi}}, \bibinfo {author} {\bibfnamefont {S.}~\bibnamefont
  {Capponi}}, \bibinfo {author} {\bibfnamefont {M.}~\bibnamefont {Mambrini}},\
  and\ \bibinfo {author} {\bibfnamefont {A.~M.}\ \bibnamefont {L{\"a}uchli}},\
  }\bibfield  {title} {\bibinfo {title} {Phase diagram of a frustrated quantum
  antiferromagnet on the honeycomb lattice: {{Magnetic}} order versus
  valence-bond crystal formation},\ }\href
  {https://doi.org/10.1103/PhysRevB.84.024406} {\bibfield  {journal} {\bibinfo
  {journal} {Phys. Rev. B}\ }\textbf {\bibinfo {volume} {84}},\ \bibinfo
  {pages} {024406} (\bibinfo {year} {2011})}\BibitemShut {NoStop}%
\bibitem [{\citenamefont {Reuther}\ \emph
  {et~al.}(2011{\natexlab{b}})\citenamefont {Reuther}, \citenamefont {Abanin},\
  and\ \citenamefont {Thomale}}]{Reuther2011a}%
  \BibitemOpen
  \bibfield  {author} {\bibinfo {author} {\bibfnamefont {J.}~\bibnamefont
  {Reuther}}, \bibinfo {author} {\bibfnamefont {D.~A.}\ \bibnamefont
  {Abanin}},\ and\ \bibinfo {author} {\bibfnamefont {R.}~\bibnamefont
  {Thomale}},\ }\bibfield  {title} {\bibinfo {title} {Magnetic order and
  paramagnetic phases in the quantum
  {{J}}{\textsubscript{1}}-{{J}}{\textsubscript{2}}-{{J}}{\textsubscript{3}}
  honeycomb model},\ }\href {https://doi.org/10.1103/PhysRevB.84.014417}
  {\bibfield  {journal} {\bibinfo  {journal} {Phys. Rev. B}\ }\textbf {\bibinfo
  {volume} {84}},\ \bibinfo {pages} {014417} (\bibinfo {year}
  {2011}{\natexlab{b}})}\BibitemShut {NoStop}%
\bibitem [{\citenamefont {Oitmaa}\ and\ \citenamefont
  {Singh}(2011)}]{Oitmaa2011}%
  \BibitemOpen
  \bibfield  {author} {\bibinfo {author} {\bibfnamefont {J.}~\bibnamefont
  {Oitmaa}}\ and\ \bibinfo {author} {\bibfnamefont {R.~R.~P.}\ \bibnamefont
  {Singh}},\ }\bibfield  {title} {\bibinfo {title} {Phase diagram of the
  {{J}}{\textsubscript{1}}-{{J}}{\textsubscript{2}}-{{J}}{\textsubscript{3}}
  {{Heisenberg}} model on the honeycomb lattice: {{A}} series expansion
  study},\ }\href {https://doi.org/10.1103/PhysRevB.84.094424} {\bibfield
  {journal} {\bibinfo  {journal} {Phys. Rev. B}\ }\textbf {\bibinfo {volume}
  {84}},\ \bibinfo {pages} {094424} (\bibinfo {year} {2011})}\BibitemShut
  {NoStop}%
\bibitem [{\citenamefont {Clark}\ \emph {et~al.}(2011)\citenamefont {Clark},
  \citenamefont {Abanin},\ and\ \citenamefont {Sondhi}}]{Clark2011}%
  \BibitemOpen
  \bibfield  {author} {\bibinfo {author} {\bibfnamefont {B.~K.}\ \bibnamefont
  {Clark}}, \bibinfo {author} {\bibfnamefont {D.~A.}\ \bibnamefont {Abanin}},\
  and\ \bibinfo {author} {\bibfnamefont {S.~L.}\ \bibnamefont {Sondhi}},\
  }\bibfield  {title} {\bibinfo {title} {Nature of the {{Spin Liquid State}} of
  the {{Hubbard Model}} on a {{Honeycomb Lattice}}},\ }\href
  {https://doi.org/10.1103/PhysRevLett.107.087204} {\bibfield  {journal}
  {\bibinfo  {journal} {Phys. Rev. Lett.}\ }\textbf {\bibinfo {volume} {107}},\
  \bibinfo {pages} {087204} (\bibinfo {year} {2011})}\BibitemShut {NoStop}%
\bibitem [{\citenamefont {Bishop}\ \emph {et~al.}(2012)\citenamefont {Bishop},
  \citenamefont {Li}, \citenamefont {Farnell},\ and\ \citenamefont
  {Campbell}}]{Bishop2012}%
  \BibitemOpen
  \bibfield  {author} {\bibinfo {author} {\bibfnamefont {R.~F.}\ \bibnamefont
  {Bishop}}, \bibinfo {author} {\bibfnamefont {P.~H.~Y.}\ \bibnamefont {Li}},
  \bibinfo {author} {\bibfnamefont {D.~J.~J.}\ \bibnamefont {Farnell}},\ and\
  \bibinfo {author} {\bibfnamefont {C.~E.}\ \bibnamefont {Campbell}},\
  }\bibfield  {title} {\bibinfo {title} {The frustrated {{Heisenberg}}
  antiferromagnet on the honeycomb lattice:
  {{J}}{\textsubscript{1}}-{{J}}{\textsubscript{2}} model},\ }\href
  {https://doi.org/10.1088/0953-8984/24/23/236002} {\bibfield  {journal}
  {\bibinfo  {journal} {J. Phys.: Condens. Matter}\ }\textbf {\bibinfo {volume}
  {24}},\ \bibinfo {pages} {236002} (\bibinfo {year} {2012})}\BibitemShut
  {NoStop}%
\bibitem [{\citenamefont {Bishop}\ \emph {et~al.}(2013)\citenamefont {Bishop},
  \citenamefont {Li},\ and\ \citenamefont {Campbell}}]{Bishop2013b}%
  \BibitemOpen
  \bibfield  {author} {\bibinfo {author} {\bibfnamefont {R.~F.}\ \bibnamefont
  {Bishop}}, \bibinfo {author} {\bibfnamefont {P.~H.~Y.}\ \bibnamefont {Li}},\
  and\ \bibinfo {author} {\bibfnamefont {C.~E.}\ \bibnamefont {Campbell}},\
  }\bibfield  {title} {\bibinfo {title} {Valence-bond crystalline order in the
  {{{\emph{s}}}} = 1/2
  {{{\emph{J}}}}{\textsubscript{1}}-{{{\emph{J}}}}{\textsubscript{2}} model on
  the honeycomb lattice},\ }\href
  {https://doi.org/10.1088/0953-8984/25/30/306002} {\bibfield  {journal}
  {\bibinfo  {journal} {J. Phys.: Condens. Matter}\ }\textbf {\bibinfo {volume}
  {25}},\ \bibinfo {pages} {306002} (\bibinfo {year} {2013})}\BibitemShut
  {NoStop}%
\bibitem [{\citenamefont {Bishop}\ \emph {et~al.}(2015)\citenamefont {Bishop},
  \citenamefont {Li}, \citenamefont {G{\"o}tze}, \citenamefont {Richter},\ and\
  \citenamefont {Campbell}}]{Bishop2015}%
  \BibitemOpen
  \bibfield  {author} {\bibinfo {author} {\bibfnamefont {R.~F.}\ \bibnamefont
  {Bishop}}, \bibinfo {author} {\bibfnamefont {P.~H.~Y.}\ \bibnamefont {Li}},
  \bibinfo {author} {\bibfnamefont {O.}~\bibnamefont {G{\"o}tze}}, \bibinfo
  {author} {\bibfnamefont {J.}~\bibnamefont {Richter}},\ and\ \bibinfo {author}
  {\bibfnamefont {C.~E.}\ \bibnamefont {Campbell}},\ }\bibfield  {title}
  {\bibinfo {title} {Frustrated {{Heisenberg}} antiferromagnet on the honeycomb
  lattice: {{Spin}} gap and low-energy parameters},\ }\href
  {https://doi.org/10.1103/PhysRevB.92.224434} {\bibfield  {journal} {\bibinfo
  {journal} {Phys. Rev. B}\ }\textbf {\bibinfo {volume} {92}},\ \bibinfo
  {pages} {224434} (\bibinfo {year} {2015})}\BibitemShut {NoStop}%
\bibitem [{\citenamefont {Sahoo}\ and\ \citenamefont
  {Flint}(2020)}]{Sahoo2020}%
  \BibitemOpen
  \bibfield  {author} {\bibinfo {author} {\bibfnamefont {J.}~\bibnamefont
  {Sahoo}}\ and\ \bibinfo {author} {\bibfnamefont {R.}~\bibnamefont {Flint}},\
  }\bibfield  {title} {\bibinfo {title} {Symmetric spin liquids on the stuffed
  honeycomb lattice},\ }\href {https://doi.org/10.1103/PhysRevB.101.115103}
  {\bibfield  {journal} {\bibinfo  {journal} {Phys. Rev. B}\ }\textbf {\bibinfo
  {volume} {101}},\ \bibinfo {pages} {115103} (\bibinfo {year}
  {2020})}\BibitemShut {NoStop}%
\bibitem [{\citenamefont {Dong}\ and\ \citenamefont {Sheng}(2020)}]{Dong2020}%
  \BibitemOpen
  \bibfield  {author} {\bibinfo {author} {\bibfnamefont {X.-Y.}\ \bibnamefont
  {Dong}}\ and\ \bibinfo {author} {\bibfnamefont {D.~N.}\ \bibnamefont
  {Sheng}},\ }\bibfield  {title} {\bibinfo {title} {Spin-1
  {{Kitaev-Heisenberg}} model on a honeycomb lattice},\ }\href
  {https://doi.org/10.1103/PhysRevB.102.121102} {\bibfield  {journal} {\bibinfo
   {journal} {Phys. Rev. B}\ }\textbf {\bibinfo {volume} {102}},\ \bibinfo
  {pages} {121102} (\bibinfo {year} {2020})}\BibitemShut {NoStop}%
\bibitem [{\citenamefont {Jin}\ \emph {et~al.}(2022)\citenamefont {Jin},
  \citenamefont {Natori}, \citenamefont {Pollmann},\ and\ \citenamefont
  {Knolle}}]{Jin2022}%
  \BibitemOpen
  \bibfield  {author} {\bibinfo {author} {\bibfnamefont {H.-K.}\ \bibnamefont
  {Jin}}, \bibinfo {author} {\bibfnamefont {W.~M.~H.}\ \bibnamefont {Natori}},
  \bibinfo {author} {\bibfnamefont {F.}~\bibnamefont {Pollmann}},\ and\
  \bibinfo {author} {\bibfnamefont {J.}~\bibnamefont {Knolle}},\ }\bibfield
  {title} {\bibinfo {title} {Unveiling the {{S}}=3/2 {{Kitaev}} honeycomb spin
  liquids},\ }\href {https://doi.org/10.1038/s41467-022-31503-0} {\bibfield
  {journal} {\bibinfo  {journal} {Nat Commun}\ }\textbf {\bibinfo {volume}
  {13}},\ \bibinfo {pages} {3813} (\bibinfo {year} {2022})}\BibitemShut
  {NoStop}%
\bibitem [{\citenamefont {Sangeetha}\ \emph {et~al.}(2021)\citenamefont
  {Sangeetha}, \citenamefont {Pakhira}, \citenamefont {Ding}, \citenamefont
  {Krause}, \citenamefont {Lee}, \citenamefont {Smetana}, \citenamefont
  {Mudring}, \citenamefont {Iversen}, \citenamefont {Furukawa},\ and\
  \citenamefont {Johnston}}]{Sangeetha2021}%
  \BibitemOpen
  \bibfield  {author} {\bibinfo {author} {\bibfnamefont {N.~S.}\ \bibnamefont
  {Sangeetha}}, \bibinfo {author} {\bibfnamefont {S.}~\bibnamefont {Pakhira}},
  \bibinfo {author} {\bibfnamefont {Q.-P.}\ \bibnamefont {Ding}}, \bibinfo
  {author} {\bibfnamefont {L.}~\bibnamefont {Krause}}, \bibinfo {author}
  {\bibfnamefont {H.-C.}\ \bibnamefont {Lee}}, \bibinfo {author} {\bibfnamefont
  {V.}~\bibnamefont {Smetana}}, \bibinfo {author} {\bibfnamefont {A.-V.}\
  \bibnamefont {Mudring}}, \bibinfo {author} {\bibfnamefont {B.~B.}\
  \bibnamefont {Iversen}}, \bibinfo {author} {\bibfnamefont {Y.}~\bibnamefont
  {Furukawa}},\ and\ \bibinfo {author} {\bibfnamefont {D.~C.}\ \bibnamefont
  {Johnston}},\ }\bibfield  {title} {\bibinfo {title} {First-order
  antiferromagnetic transitions of
  {{SrMn}}{\textsubscript{2}}{{P}}{\textsubscript{2}} and
  {{CaMn}}{\textsubscript{2}}{{P}}{\textsubscript{2}} single crystals
  containing corrugated-honeycomb {{Mn}} sublattices},\ }\bibfield  {journal}
  {\bibinfo  {journal} {Proc. Natl. Acad. Sci.}\ }\textbf {\bibinfo {volume}
  {118}},\ \href {https://doi.org/10.1073/pnas.2108724118}
  {10.1073/pnas.2108724118} (\bibinfo {year} {2021})\BibitemShut {NoStop}%
\bibitem [{\citenamefont {Sangeetha}\ \emph {et~al.}(2016)\citenamefont
  {Sangeetha}, \citenamefont {Pandey}, \citenamefont {Benson},\ and\
  \citenamefont {Johnston}}]{Sangeetha2016}%
  \BibitemOpen
  \bibfield  {author} {\bibinfo {author} {\bibfnamefont {N.~S.}\ \bibnamefont
  {Sangeetha}}, \bibinfo {author} {\bibfnamefont {A.}~\bibnamefont {Pandey}},
  \bibinfo {author} {\bibfnamefont {Z.~A.}\ \bibnamefont {Benson}},\ and\
  \bibinfo {author} {\bibfnamefont {D.~C.}\ \bibnamefont {Johnston}},\
  }\bibfield  {title} {\bibinfo {title} {Strong magnetic correlations to 900
  {{K}} in single crystals of the trigonal antiferromagnetic insulators
  {{SrMn}}{\textsubscript{2}}{{As}}{\textsubscript{2}} and
  {{CaMn}}{\textsubscript{2}}{{As}}{\textsubscript{2}}},\ }\href
  {https://doi.org/10.1103/PhysRevB.94.094417} {\bibfield  {journal} {\bibinfo
  {journal} {Phys. Rev. B}\ }\textbf {\bibinfo {volume} {94}},\ \bibinfo
  {pages} {094417} (\bibinfo {year} {2016})}\BibitemShut {NoStop}%
\bibitem [{\citenamefont {Simonson}\ \emph {et~al.}(2012)\citenamefont
  {Simonson}, \citenamefont {Smith}, \citenamefont {Post}, \citenamefont
  {Pezzoli}, \citenamefont {{Kistner-Morris}}, \citenamefont {McNally},
  \citenamefont {Hassinger}, \citenamefont {Nelson}, \citenamefont {Kotliar},
  \citenamefont {Basov},\ and\ \citenamefont {Aronson}}]{Simonson2012a}%
  \BibitemOpen
  \bibfield  {author} {\bibinfo {author} {\bibfnamefont {J.~W.}\ \bibnamefont
  {Simonson}}, \bibinfo {author} {\bibfnamefont {G.~J.}\ \bibnamefont {Smith}},
  \bibinfo {author} {\bibfnamefont {K.}~\bibnamefont {Post}}, \bibinfo {author}
  {\bibfnamefont {M.}~\bibnamefont {Pezzoli}}, \bibinfo {author} {\bibfnamefont
  {J.~J.}\ \bibnamefont {{Kistner-Morris}}}, \bibinfo {author} {\bibfnamefont
  {D.~E.}\ \bibnamefont {McNally}}, \bibinfo {author} {\bibfnamefont {J.~E.}\
  \bibnamefont {Hassinger}}, \bibinfo {author} {\bibfnamefont {C.~S.}\
  \bibnamefont {Nelson}}, \bibinfo {author} {\bibfnamefont {G.}~\bibnamefont
  {Kotliar}}, \bibinfo {author} {\bibfnamefont {D.~N.}\ \bibnamefont {Basov}},\
  and\ \bibinfo {author} {\bibfnamefont {M.~C.}\ \bibnamefont {Aronson}},\
  }\bibfield  {title} {\bibinfo {title} {Magnetic and structural phase diagram
  of {{CaMn}}{\textsubscript{2}}{{Sb}}{\textsubscript{2}}},\ }\href
  {https://doi.org/10.1103/PhysRevB.86.184430} {\bibfield  {journal} {\bibinfo
  {journal} {Phys. Rev. B}\ }\textbf {\bibinfo {volume} {86}},\ \bibinfo
  {pages} {184430} (\bibinfo {year} {2012})}\BibitemShut {NoStop}%
\bibitem [{\citenamefont {Sangeetha}\ \emph {et~al.}(2018)\citenamefont
  {Sangeetha}, \citenamefont {Smetana}, \citenamefont {Mudring},\ and\
  \citenamefont {Johnston}}]{Sangeetha2018}%
  \BibitemOpen
  \bibfield  {author} {\bibinfo {author} {\bibfnamefont {N.~S.}\ \bibnamefont
  {Sangeetha}}, \bibinfo {author} {\bibfnamefont {V.}~\bibnamefont {Smetana}},
  \bibinfo {author} {\bibfnamefont {A.-V.}\ \bibnamefont {Mudring}},\ and\
  \bibinfo {author} {\bibfnamefont {D.~C.}\ \bibnamefont {Johnston}},\
  }\bibfield  {title} {\bibinfo {title} {Antiferromagnetism in semiconducting
  {{SrMn}}{$_{2}$}{{Sb}}{$_2$} and {{BaMn}}{$_{2}$}{{Sb}}{$_2$} single
  crystals},\ }\href {https://doi.org/10.1103/PhysRevB.97.014402} {\bibfield
  {journal} {\bibinfo  {journal} {Phys. Rev. B}\ }\textbf {\bibinfo {volume}
  {97}},\ \bibinfo {pages} {014402} (\bibinfo {year} {2018})}\BibitemShut
  {NoStop}%
\bibitem [{\citenamefont {Das}\ \emph {et~al.}(2017)\citenamefont {Das},
  \citenamefont {Sangeetha}, \citenamefont {Pandey}, \citenamefont {Benson},
  \citenamefont {Heitmann}, \citenamefont {Johnston}, \citenamefont {Goldman},\
  and\ \citenamefont {Kreyssig}}]{Das2017}%
  \BibitemOpen
  \bibfield  {author} {\bibinfo {author} {\bibfnamefont {P.}~\bibnamefont
  {Das}}, \bibinfo {author} {\bibfnamefont {N.~S.}\ \bibnamefont {Sangeetha}},
  \bibinfo {author} {\bibfnamefont {A.}~\bibnamefont {Pandey}}, \bibinfo
  {author} {\bibfnamefont {Z.~A.}\ \bibnamefont {Benson}}, \bibinfo {author}
  {\bibfnamefont {T.~W.}\ \bibnamefont {Heitmann}}, \bibinfo {author}
  {\bibfnamefont {D.~C.}\ \bibnamefont {Johnston}}, \bibinfo {author}
  {\bibfnamefont {A.~I.}\ \bibnamefont {Goldman}},\ and\ \bibinfo {author}
  {\bibfnamefont {A.}~\bibnamefont {Kreyssig}},\ }\bibfield  {title} {\bibinfo
  {title} {Collinear antiferromagnetism in trigonal
  {{SrMn}}{\textsubscript{2}}{{As}}{\textsubscript{2}} revealed by
  single-crystal neutron diffraction},\ }\href
  {https://doi.org/10.1088/0953-8984/29/3/035802} {\bibfield  {journal}
  {\bibinfo  {journal} {J. Phys.: Condens. Matter}\ }\textbf {\bibinfo {volume}
  {29}},\ \bibinfo {pages} {035802} (\bibinfo {year} {2017})}\BibitemShut
  {NoStop}%
\bibitem [{\citenamefont {Bridges}\ \emph {et~al.}(2009)\citenamefont
  {Bridges}, \citenamefont {Krishnamurthy}, \citenamefont {Poulton},
  \citenamefont {Paranthaman}, \citenamefont {Sales}, \citenamefont {Myers},\
  and\ \citenamefont {Bobev}}]{Bridges2009}%
  \BibitemOpen
  \bibfield  {author} {\bibinfo {author} {\bibfnamefont {C.}~\bibnamefont
  {Bridges}}, \bibinfo {author} {\bibfnamefont {V.}~\bibnamefont
  {Krishnamurthy}}, \bibinfo {author} {\bibfnamefont {S.}~\bibnamefont
  {Poulton}}, \bibinfo {author} {\bibfnamefont {M.}~\bibnamefont
  {Paranthaman}}, \bibinfo {author} {\bibfnamefont {B.}~\bibnamefont {Sales}},
  \bibinfo {author} {\bibfnamefont {C.}~\bibnamefont {Myers}},\ and\ \bibinfo
  {author} {\bibfnamefont {S.}~\bibnamefont {Bobev}},\ }\bibfield  {title}
  {\bibinfo {title} {Magnetic order in
  {{CaMn}}{\textsubscript{2}}{{Sb}}{\textsubscript{2}} studied via powder
  neutron diffraction},\ }\href {https://doi.org/10.1016/j.jmmm.2009.07.015}
  {\bibfield  {journal} {\bibinfo  {journal} {Journal of Magnetism and Magnetic
  Materials}\ }\textbf {\bibinfo {volume} {321}},\ \bibinfo {pages} {3653}
  (\bibinfo {year} {2009})}\BibitemShut {NoStop}%
\bibitem [{\citenamefont {Gibson}\ \emph {et~al.}(2015)\citenamefont {Gibson},
  \citenamefont {Wu}, \citenamefont {Liang}, \citenamefont {Ali}, \citenamefont
  {Ong}, \citenamefont {Huang},\ and\ \citenamefont {Cava}}]{Gibson2015}%
  \BibitemOpen
  \bibfield  {author} {\bibinfo {author} {\bibfnamefont {Q.~D.}\ \bibnamefont
  {Gibson}}, \bibinfo {author} {\bibfnamefont {H.}~\bibnamefont {Wu}}, \bibinfo
  {author} {\bibfnamefont {T.}~\bibnamefont {Liang}}, \bibinfo {author}
  {\bibfnamefont {M.~N.}\ \bibnamefont {Ali}}, \bibinfo {author} {\bibfnamefont
  {N.~P.}\ \bibnamefont {Ong}}, \bibinfo {author} {\bibfnamefont
  {Q.}~\bibnamefont {Huang}},\ and\ \bibinfo {author} {\bibfnamefont {R.~J.}\
  \bibnamefont {Cava}},\ }\bibfield  {title} {\bibinfo {title} {Magnetic and
  electronic properties of
  {{CaMn}}{\textsubscript{2}}{{Bi}}{\textsubscript{2}}: {{A}} possible
  hybridization gap semiconductor},\ }\href
  {https://doi.org/10.1103/PhysRevB.91.085128} {\bibfield  {journal} {\bibinfo
  {journal} {Phys. Rev. B}\ }\textbf {\bibinfo {volume} {91}},\ \bibinfo
  {pages} {085128} (\bibinfo {year} {2015})}\BibitemShut {NoStop}%
\bibitem [{\citenamefont {Ratcliff~II}\ \emph {et~al.}(2009)\citenamefont
  {Ratcliff~II}, \citenamefont {Lima~Sharma}, \citenamefont {Gomes},
  \citenamefont {Gonzalez}, \citenamefont {Huang},\ and\ \citenamefont
  {Singleton}}]{Ratcliff2009}%
  \BibitemOpen
  \bibfield  {author} {\bibinfo {author} {\bibfnamefont {W.}~\bibnamefont
  {Ratcliff~II}}, \bibinfo {author} {\bibfnamefont {A.}~\bibnamefont
  {Lima~Sharma}}, \bibinfo {author} {\bibfnamefont {A.}~\bibnamefont {Gomes}},
  \bibinfo {author} {\bibfnamefont {J.}~\bibnamefont {Gonzalez}}, \bibinfo
  {author} {\bibfnamefont {Q.}~\bibnamefont {Huang}},\ and\ \bibinfo {author}
  {\bibfnamefont {J.}~\bibnamefont {Singleton}},\ }\bibfield  {title} {\bibinfo
  {title} {The magnetic ground state of
  {{CaMn}}{\textsubscript{2}}{{Sb}}{\textsubscript{2}}},\ }\href
  {https://doi.org/10.1016/j.jmmm.2009.03.054} {\bibfield  {journal} {\bibinfo
  {journal} {J. Magn. Magn. Mater.}\ }\textbf {\bibinfo {volume} {321}},\
  \bibinfo {pages} {2612} (\bibinfo {year} {2009})}\BibitemShut {NoStop}%
\bibitem [{\citenamefont {Islam}\ \emph {et~al.}(2020)\citenamefont {Islam},
  \citenamefont {Gordon}, \citenamefont {Das}, \citenamefont {Liu},
  \citenamefont {Ke}, \citenamefont {Abernathy}, \citenamefont {McQueeney},\
  and\ \citenamefont {Vaknin}}]{Islam2020a}%
  \BibitemOpen
  \bibfield  {author} {\bibinfo {author} {\bibfnamefont {F.}~\bibnamefont
  {Islam}}, \bibinfo {author} {\bibfnamefont {E.}~\bibnamefont {Gordon}},
  \bibinfo {author} {\bibfnamefont {P.}~\bibnamefont {Das}}, \bibinfo {author}
  {\bibfnamefont {Y.}~\bibnamefont {Liu}}, \bibinfo {author} {\bibfnamefont
  {L.}~\bibnamefont {Ke}}, \bibinfo {author} {\bibfnamefont {D.~L.}\
  \bibnamefont {Abernathy}}, \bibinfo {author} {\bibfnamefont {R.~J.}\
  \bibnamefont {McQueeney}},\ and\ \bibinfo {author} {\bibfnamefont
  {D.}~\bibnamefont {Vaknin}},\ }\bibfield  {title} {\bibinfo {title} {Spin
  dynamics in antiferromagnetic oxypnictides and fluoropnictides: {{LaMnAsO}},
  {{LaMnSbO}}, and {{BaMnAsF}}},\ }\href
  {https://doi.org/10.1103/PhysRevB.101.155119} {\bibfield  {journal} {\bibinfo
   {journal} {Phys. Rev. B}\ }\textbf {\bibinfo {volume} {101}},\ \bibinfo
  {pages} {155119} (\bibinfo {year} {2020})}\BibitemShut {NoStop}%
\bibitem [{\citenamefont {Brock}\ \emph {et~al.}(1994)\citenamefont {Brock},
  \citenamefont {Greedan},\ and\ \citenamefont {Kauzlarich}}]{Brock1994}%
  \BibitemOpen
  \bibfield  {author} {\bibinfo {author} {\bibfnamefont {S.~L.}\ \bibnamefont
  {Brock}}, \bibinfo {author} {\bibfnamefont {J.}~\bibnamefont {Greedan}},\
  and\ \bibinfo {author} {\bibfnamefont {S.~M.}\ \bibnamefont {Kauzlarich}},\
  }\bibfield  {title} {\bibinfo {title} {Resistivity and magnetism of
  {{AMn}}{\textsubscript{2}}{{P}}{\textsubscript{2}} ({{A}} = {{Sr}}, {{Ba}}):
  {{The}} effect of structure type on physical properties},\ }\href
  {https://doi.org/10.1006/jssc.1994.1375} {\bibfield  {journal} {\bibinfo
  {journal} {J. Solid State Chem.}\ }\textbf {\bibinfo {volume} {113}},\
  \bibinfo {pages} {303} (\bibinfo {year} {1994})}\BibitemShut {NoStop}%
\bibitem [{\citenamefont {Little}\ \emph {et~al.}(2020)\citenamefont {Little},
  \citenamefont {Lee}, \citenamefont {John}, \citenamefont {Doyle},
  \citenamefont {Maniv}, \citenamefont {Nair}, \citenamefont {Chen},
  \citenamefont {Rees}, \citenamefont {Venderbos}, \citenamefont {Fernandes},
  \citenamefont {Analytis},\ and\ \citenamefont {Orenstein}}]{Little2020}%
  \BibitemOpen
  \bibfield  {author} {\bibinfo {author} {\bibfnamefont {A.}~\bibnamefont
  {Little}}, \bibinfo {author} {\bibfnamefont {C.}~\bibnamefont {Lee}},
  \bibinfo {author} {\bibfnamefont {C.}~\bibnamefont {John}}, \bibinfo {author}
  {\bibfnamefont {S.}~\bibnamefont {Doyle}}, \bibinfo {author} {\bibfnamefont
  {E.}~\bibnamefont {Maniv}}, \bibinfo {author} {\bibfnamefont {N.~L.}\
  \bibnamefont {Nair}}, \bibinfo {author} {\bibfnamefont {W.}~\bibnamefont
  {Chen}}, \bibinfo {author} {\bibfnamefont {D.}~\bibnamefont {Rees}}, \bibinfo
  {author} {\bibfnamefont {J.~W.~F.}\ \bibnamefont {Venderbos}}, \bibinfo
  {author} {\bibfnamefont {R.~M.}\ \bibnamefont {Fernandes}}, \bibinfo {author}
  {\bibfnamefont {J.~G.}\ \bibnamefont {Analytis}},\ and\ \bibinfo {author}
  {\bibfnamefont {J.}~\bibnamefont {Orenstein}},\ }\bibfield  {title} {\bibinfo
  {title} {Three-state nematicity in the triangular lattice antiferromagnet
  {{Fe}}{\textsubscript{1/3}}{{NbS}}{\textsubscript{2}}},\ }\href
  {https://doi.org/10.1038/s41563-020-0681-0} {\bibfield  {journal} {\bibinfo
  {journal} {Nat. Mater.}\ }\textbf {\bibinfo {volume} {19}},\ \bibinfo {pages}
  {1062} (\bibinfo {year} {2020})}\BibitemShut {NoStop}%
\bibitem [{\citenamefont {Fernandes}\ \emph {et~al.}(2019)\citenamefont
  {Fernandes}, \citenamefont {Orth},\ and\ \citenamefont
  {Schmalian}}]{Fernandes2019}%
  \BibitemOpen
  \bibfield  {author} {\bibinfo {author} {\bibfnamefont {R.~M.}\ \bibnamefont
  {Fernandes}}, \bibinfo {author} {\bibfnamefont {P.~P.}\ \bibnamefont
  {Orth}},\ and\ \bibinfo {author} {\bibfnamefont {J.}~\bibnamefont
  {Schmalian}},\ }\bibfield  {title} {\bibinfo {title} {Intertwined {{Vestigial
  Order}} in {{Quantum Materials}}: {{Nematicity}} and {{Beyond}}},\ }\href
  {https://doi.org/10.1146/annurev-conmatphys-031218-013200} {\bibfield
  {journal} {\bibinfo  {journal} {Annu. Rev. Condens. Matter Phys.}\ }\textbf
  {\bibinfo {volume} {10}},\ \bibinfo {pages} {133} (\bibinfo {year}
  {2019})}\BibitemShut {NoStop}%
\bibitem [{\citenamefont {McNally}\ \emph {et~al.}(2015)\citenamefont
  {McNally}, \citenamefont {Simonson}, \citenamefont {{Kistner-Morris}},
  \citenamefont {Smith}, \citenamefont {Hassinger}, \citenamefont
  {{DeBeer-Schmitt}}, \citenamefont {Kolesnikov}, \citenamefont {Zaliznyak},\
  and\ \citenamefont {Aronson}}]{Mcnally2015}%
  \BibitemOpen
  \bibfield  {author} {\bibinfo {author} {\bibfnamefont {D.~E.}\ \bibnamefont
  {McNally}}, \bibinfo {author} {\bibfnamefont {J.~W.}\ \bibnamefont
  {Simonson}}, \bibinfo {author} {\bibfnamefont {J.~J.}\ \bibnamefont
  {{Kistner-Morris}}}, \bibinfo {author} {\bibfnamefont {G.~J.}\ \bibnamefont
  {Smith}}, \bibinfo {author} {\bibfnamefont {J.~E.}\ \bibnamefont
  {Hassinger}}, \bibinfo {author} {\bibfnamefont {L.}~\bibnamefont
  {{DeBeer-Schmitt}}}, \bibinfo {author} {\bibfnamefont {A.~I.}\ \bibnamefont
  {Kolesnikov}}, \bibinfo {author} {\bibfnamefont {I.~A.}\ \bibnamefont
  {Zaliznyak}},\ and\ \bibinfo {author} {\bibfnamefont {M.~C.}\ \bibnamefont
  {Aronson}},\ }\bibfield  {title} {\bibinfo {title}
  {{{CaMn}}{\textsubscript{2}}{{Sb}}{\textsubscript{2}}: {{Spin}} waves on a
  frustrated antiferromagnetic honeycomb lattice},\ }\href
  {https://doi.org/10.1103/PhysRevB.91.180407} {\bibfield  {journal} {\bibinfo
  {journal} {Phys. Rev. B}\ }\textbf {\bibinfo {volume} {91}},\ \bibinfo
  {pages} {180407} (\bibinfo {year} {2015})}\BibitemShut {NoStop}%
\bibitem [{SM-()}]{SM-CaMn2P2}%
  \BibitemOpen
  \bibfield  {title} {\bibinfo {title} {See {{Supplemental Material}} at
  {{http://link.aps.org/supplemental/}} for details on constructing the
  magnetic structure and systematically searching other magnetic structures.},\
  }\href@noop {} {\ }\BibitemShut {NoStop}%
\bibitem [{\citenamefont {{Perez-Mato}}\ \emph {et~al.}(2015)\citenamefont
  {{Perez-Mato}}, \citenamefont {Gallego}, \citenamefont {Tasci}, \citenamefont
  {Elcoro}, \citenamefont {{de la Flor}},\ and\ \citenamefont
  {Aroyo}}]{bilbao}%
  \BibitemOpen
  \bibfield  {author} {\bibinfo {author} {\bibfnamefont {J.}~\bibnamefont
  {{Perez-Mato}}}, \bibinfo {author} {\bibfnamefont {S.}~\bibnamefont
  {Gallego}}, \bibinfo {author} {\bibfnamefont {E.}~\bibnamefont {Tasci}},
  \bibinfo {author} {\bibfnamefont {L.}~\bibnamefont {Elcoro}}, \bibinfo
  {author} {\bibfnamefont {G.}~\bibnamefont {{de la Flor}}},\ and\ \bibinfo
  {author} {\bibfnamefont {M.}~\bibnamefont {Aroyo}},\ }\bibfield  {title}
  {\bibinfo {title} {Symmetry-based computational tools for magnetic
  crystallography},\ }\href
  {https://doi.org/10.1146/annurev-matsci-070214-021008} {\bibfield  {journal}
  {\bibinfo  {journal} {Annu. Rev. Mater. Res.}\ }\textbf {\bibinfo {volume}
  {45}},\ \bibinfo {pages} {217} (\bibinfo {year} {2015})}\BibitemShut
  {NoStop}%
\bibitem [{\citenamefont {Ni}\ \emph {et~al.}(2008)\citenamefont {Ni},
  \citenamefont {Nandi}, \citenamefont {Kreyssig}, \citenamefont {Goldman},
  \citenamefont {Mun}, \citenamefont {Bud'ko},\ and\ \citenamefont
  {Canfield}}]{Ni2008}%
  \BibitemOpen
  \bibfield  {author} {\bibinfo {author} {\bibfnamefont {N.}~\bibnamefont
  {Ni}}, \bibinfo {author} {\bibfnamefont {S.}~\bibnamefont {Nandi}}, \bibinfo
  {author} {\bibfnamefont {A.}~\bibnamefont {Kreyssig}}, \bibinfo {author}
  {\bibfnamefont {A.~I.}\ \bibnamefont {Goldman}}, \bibinfo {author}
  {\bibfnamefont {E.~D.}\ \bibnamefont {Mun}}, \bibinfo {author} {\bibfnamefont
  {S.~L.}\ \bibnamefont {Bud'ko}},\ and\ \bibinfo {author} {\bibfnamefont
  {P.~C.}\ \bibnamefont {Canfield}},\ }\bibfield  {title} {\bibinfo {title}
  {First-order structural phase transition in
  {{CaFe}}{\textsubscript{2}}{{As}}{\textsubscript{2}}},\ }\href
  {https://doi.org/10.1103/PhysRevB.78.014523} {\bibfield  {journal} {\bibinfo
  {journal} {Phys. Rev. B}\ }\textbf {\bibinfo {volume} {78}},\ \bibinfo
  {pages} {014523} (\bibinfo {year} {2008})}\BibitemShut {NoStop}%
\bibitem [{\citenamefont {Fernandes}\ \emph {et~al.}(2012)\citenamefont
  {Fernandes}, \citenamefont {Chubukov}, \citenamefont {Knolle}, \citenamefont
  {Eremin},\ and\ \citenamefont {Schmalian}}]{Fernandes2012}%
  \BibitemOpen
  \bibfield  {author} {\bibinfo {author} {\bibfnamefont {R.~M.}\ \bibnamefont
  {Fernandes}}, \bibinfo {author} {\bibfnamefont {A.~V.}\ \bibnamefont
  {Chubukov}}, \bibinfo {author} {\bibfnamefont {J.}~\bibnamefont {Knolle}},
  \bibinfo {author} {\bibfnamefont {I.}~\bibnamefont {Eremin}},\ and\ \bibinfo
  {author} {\bibfnamefont {J.}~\bibnamefont {Schmalian}},\ }\bibfield  {title}
  {\bibinfo {title} {Preemptive nematic order, pseudogap, and orbital order in
  the iron pnictides},\ }\href {https://doi.org/10.1103/PhysRevB.85.024534}
  {\bibfield  {journal} {\bibinfo  {journal} {Phys. Rev. B}\ }\textbf {\bibinfo
  {volume} {85}},\ \bibinfo {pages} {024534} (\bibinfo {year}
  {2012})}\BibitemShut {NoStop}%
\bibitem [{\citenamefont {Janke}\ and\ \citenamefont
  {Villanova}(1997)}]{Janke1997}%
  \BibitemOpen
  \bibfield  {author} {\bibinfo {author} {\bibfnamefont {W.}~\bibnamefont
  {Janke}}\ and\ \bibinfo {author} {\bibfnamefont {R.}~\bibnamefont
  {Villanova}},\ }\bibfield  {title} {\bibinfo {title} {Three-dimensional
  3-state {{Potts}} model revisited with new techniques},\ }\href
  {https://doi.org/10.1016/S0550-3213(96)00710-9} {\bibfield  {journal}
  {\bibinfo  {journal} {Nuclear Physics B}\ }\textbf {\bibinfo {volume}
  {489}},\ \bibinfo {pages} {679} (\bibinfo {year} {1997})}\BibitemShut
  {NoStop}%
\bibitem [{\citenamefont {{Li, Y. J.}}\ \emph {et~al.}(2020)\citenamefont {{Li,
  Y. J.}}, \citenamefont {{Jin, F.}}, \citenamefont {{Mi, Z. Y.}},
  \citenamefont {{Guo, J.}}, \citenamefont {{Wu, W.}}, \citenamefont {{Yu, Z.
  H.}}, \citenamefont {{Wu, D. S.}}, \citenamefont {{Na, S. H.}}, \citenamefont
  {{Mu, C.}}, \citenamefont {{Zhou, X. B.}}, \citenamefont {{Li, Z.}},
  \citenamefont {{Liu, K.}}, \citenamefont {{Sun, L. L.}}, \citenamefont
  {{Zhang, Q. M.}}, \citenamefont {{Xiang, T.}}, \citenamefont {{Li, G.}},\
  and\ \citenamefont {{Luo, J. L.}}}]{Li2020}%
  \BibitemOpen
  \bibfield  {author} {\bibinfo {author} {\bibnamefont {{Li, Y. J.}}}, \bibinfo
  {author} {\bibnamefont {{Jin, F.}}}, \bibinfo {author} {\bibnamefont {{Mi, Z.
  Y.}}}, \bibinfo {author} {\bibnamefont {{Guo, J.}}}, \bibinfo {author}
  {\bibnamefont {{Wu, W.}}}, \bibinfo {author} {\bibnamefont {{Yu, Z. H.}}},
  \bibinfo {author} {\bibnamefont {{Wu, D. S.}}}, \bibinfo {author}
  {\bibnamefont {{Na, S. H.}}}, \bibinfo {author} {\bibnamefont {{Mu, C.}}},
  \bibinfo {author} {\bibnamefont {{Zhou, X. B.}}}, \bibinfo {author}
  {\bibnamefont {{Li, Z.}}}, \bibinfo {author} {\bibnamefont {{Liu, K.}}},
  \bibinfo {author} {\bibnamefont {{Sun, L. L.}}}, \bibinfo {author}
  {\bibnamefont {{Zhang, Q. M.}}}, \bibinfo {author} {\bibnamefont {{Xiang,
  T.}}}, \bibinfo {author} {\bibnamefont {{Li, G.}}},\ and\ \bibinfo {author}
  {\bibnamefont {{Luo, J. L.}}},\ }\bibfield  {title} {\bibinfo {title}
  {First-order transition in trigonal structure
  {{CaMn}}{\textsubscript{2}}{{P}}{\textsubscript{2}}},\ }\href
  {https://doi.org/10.1209/0295-5075/132/46001} {\bibfield  {journal} {\bibinfo
   {journal} {EPL}\ }\textbf {\bibinfo {volume} {132}},\ \bibinfo {pages}
  {46001} (\bibinfo {year} {2020})}\BibitemShut {NoStop}%
\end{thebibliography}%

\clearpage
\pagebreak

{\center{ 
{\bf Supplemental Material \\
Frustrated Magnetic Cycloidal Structure and Emergent Potts Nematicity in CaMn$_2$P$_2$} \\~\\

{Farhan Islam$^{1,2}$, Tha\'is V. Trevisan$^{1,2}$, Thomas Heitmann$^3$, Santanu Pakhira$^1$, Simon~X.~M.~Riberolles$^1$, N.~S.~Sangeetha$^1$, David C. Johnston$^{1,2}$, Peter P. Orth$^{1,2}$, David Vaknin$^{1,2}$}\\~\\
{\it $^1$ {Ames National Laboratory, Iowa State University, Ames, Iowa 50011, USA} \\
$^2${Department of Physics and Astronomy, Iowa State University, Ames, Iowa 50011, USA} \\
$^3${The Missouri Research Reactor and Department of Physics and Astronomy, University of Missouri, Columbia, Missouri 65211, USA}\\
}
}
}

\setcounter{page}{1}
\setcounter{figure}{0}
\setcounter{equation}{0}
\setcounter{table}{0}
\setcounter{section}{0}

\renewcommand{\thefigure}{S\arabic{figure}}
\renewcommand{\theequation}{S\arabic{equation}}
\renewcommand{\thetable}{S\arabic{table}}
\renewcommand{\thepage}{S\arabic{page}}

\section{Constructing the Magnetic Structure}

\begin{figure*}[h]
\centering
\includegraphics[width=7 in]{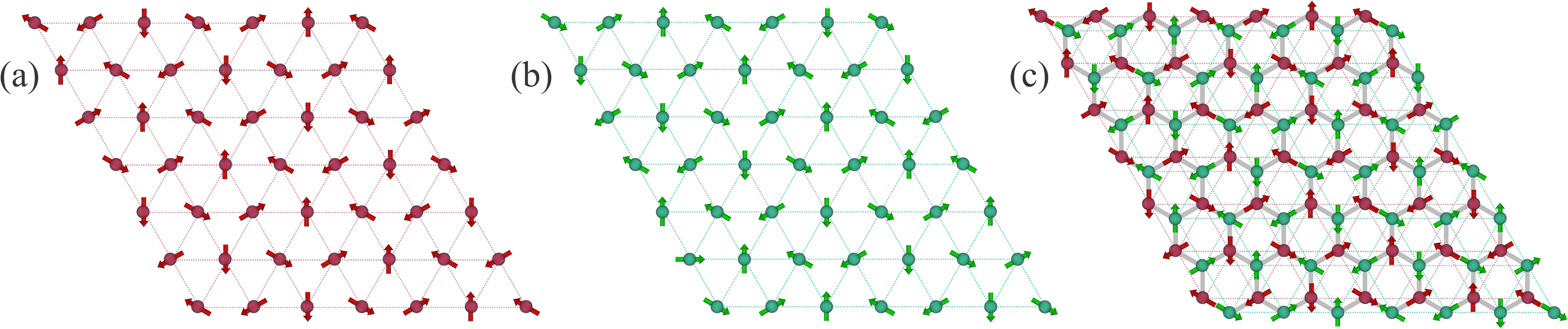}
\caption{(a) and (b) depict the $6\times6$ in-plane trigonal sublattices that form the honeycomb structure. In (a), we assign a spin along a high-symmetry direction at the origin (lower-left corner), and the nearest neighbors along the $a$- and $b$- axes are successively rotated by a $60^\circ$ angle in the clockwise direction to create the sublattice. The sublattice shown in (b) is constructed by flipping the spin direction at the origin (lower-left corner) with respect to the one in the origin of (a), and the nearest neighbors along the $a$- and $b$- axes are successively rotated by a $60^\circ$ angle in the clockwise direction. (c) The corrugated honeycomb structure with magnetic spacegroup $P_Ac$ is constructed by stacking the sublattices shown in (a) and (b).}
\label{Fig:SM:constructingStructure}
\end{figure*}

We describe the proposed magnetic structures with space group $P_Ac$, shown in Fig.~\ref{Fig:ModelCalc}(b) in the main text. The magnetic structure is constructed by generating two $6\times6$ in-plane trigonal sublattices that are stacked together to form the honeycomb structure. In Fig.~\ref{Fig:SM:constructingStructure}(a), the trigonal sublattice is constructed by assigning a spin at the origin (lower-left corner) pointed along one of the high-symmetry directions, and successively rotating the nearest neighbors in the $a$- and $b$- directions by a 60$^\circ$ angle in the counter-clockwise direction. In Fig.~\ref{Fig:SM:constructingStructure}(b) we construct the other trigonal sublattice for magnetic structure. The spin at the origin (lower-left corner) is flipped with respect to the spin at the origin of the trigonal sublattice depicted in Fig.~\ref{Fig:SM:constructingStructure}(a). The nearest neighbors along the $a-$ and $b-$ directions are rotated successively by a 60$^\circ$ angle in the counter-clockwise direction, the same way as before, to construct the other sublattice. The magnetic structure with spacegroup $P_Ac$ is formed by stacking the trigonal sublattice shown in Fig.~\ref{Fig:SM:constructingStructure}(b) on the trigonal sublattice shown in Fig.~\ref{Fig:SM:constructingStructure}(a) to form the corrugated honecycomb lattice, as depicted in Fig.~\ref{Fig:SM:constructingStructure}(c).

\section{Systematically Searching Other Magnetic Structures}

\begin{figure*}[h]
\centering
\includegraphics[width=\linewidth]{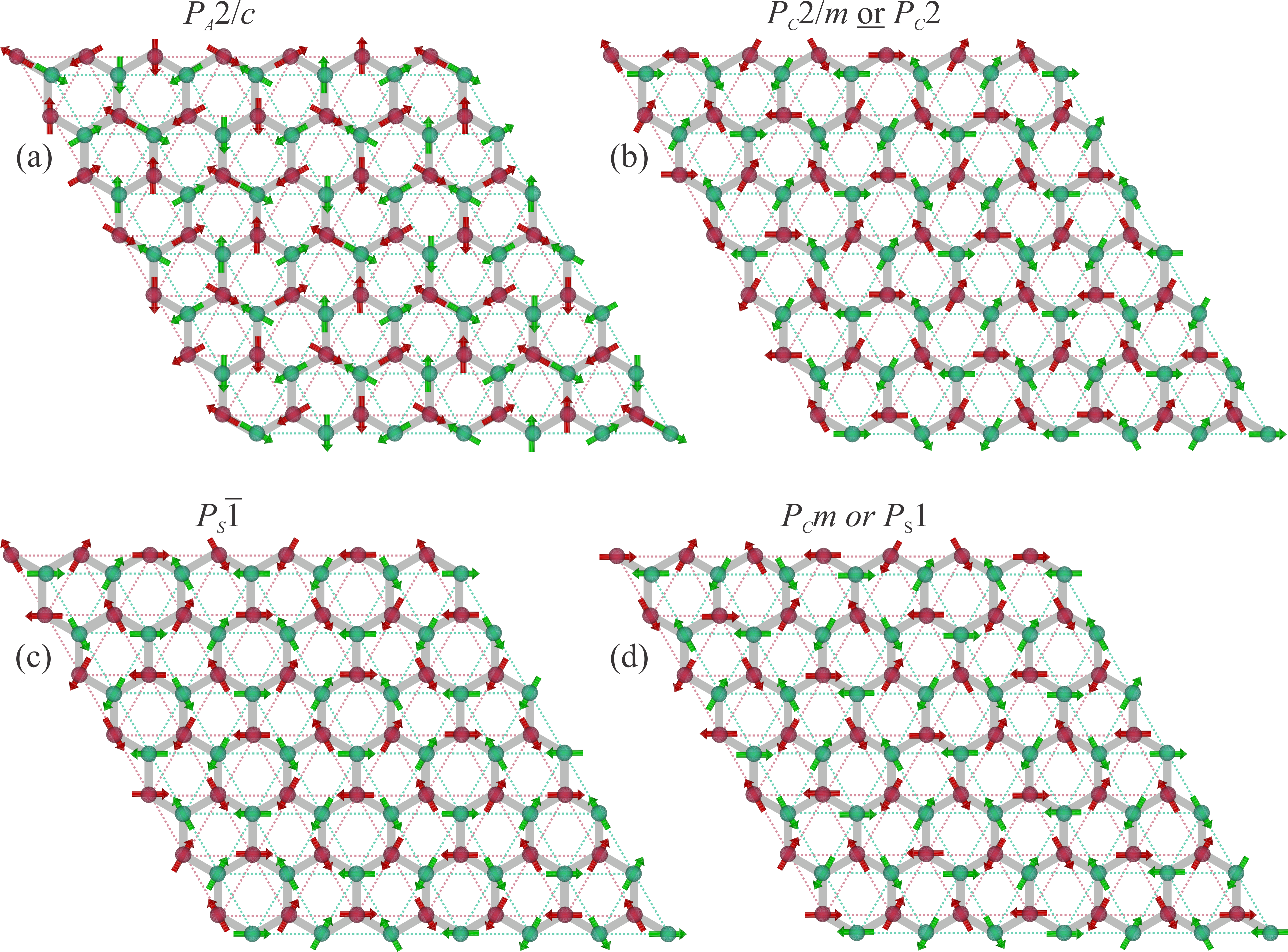}
\caption{Illustration of possible magnetic structures allowed by the Bilbao magnetic spacegroups \cite{bilbao} with calculated intensities consistent with the experimental observation. Calculated intensities for these configurations are similar to the one shown in the main text for $P_Ac$ magnetic space group.}
\label{Fig:inclMagStr}
\end{figure*}

\begin{figure*}[h]
\centering
\includegraphics[width=\linewidth]{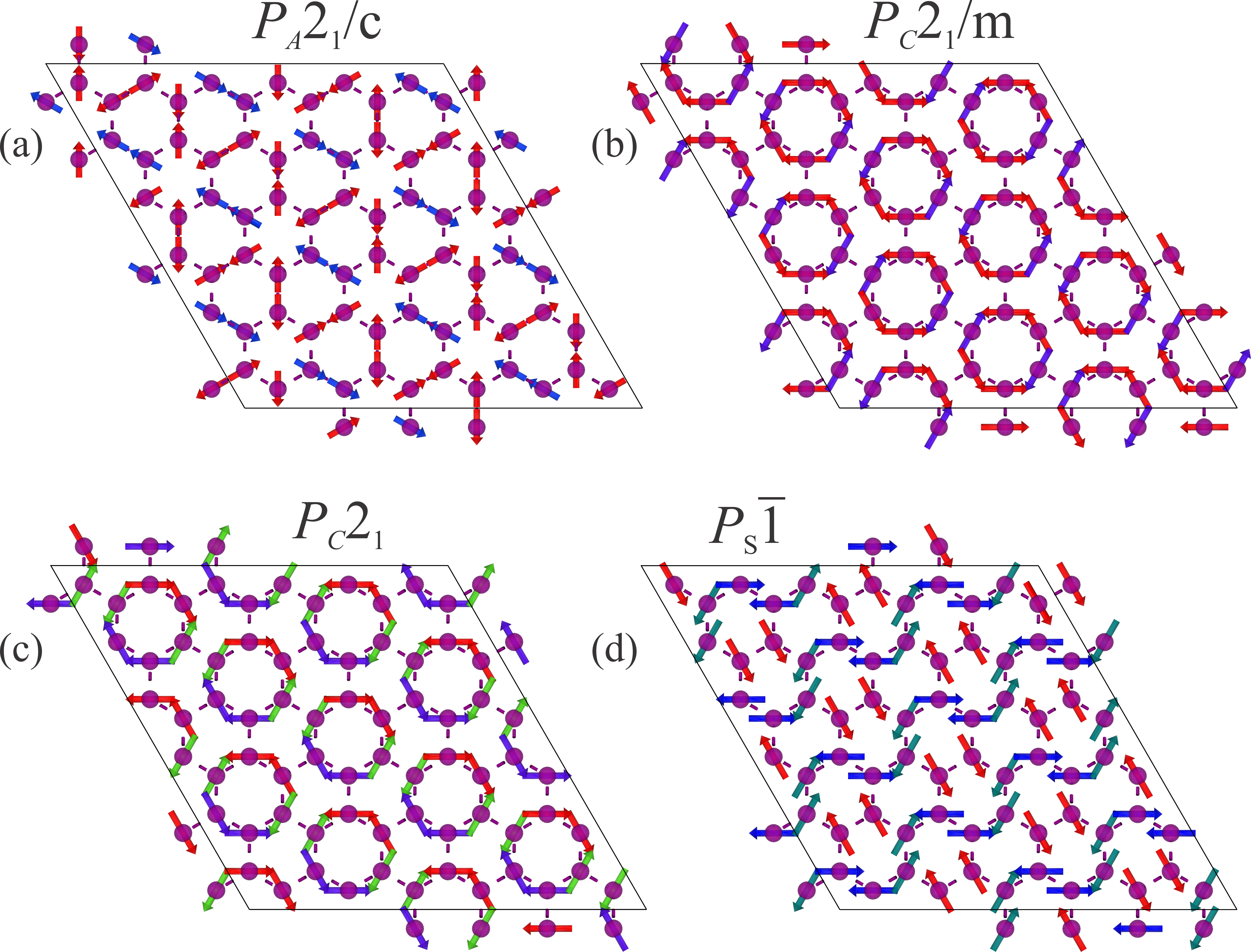}
\caption{Illustration of possible magnetic structures allowed by the Bilbao magnetic spacegroups \cite{bilbao} with calculated intensities inconsistent with the experimental observation. In particular, these configurations show intensities at ($\pm$1/6, $\pm$1/6, 0), which are not observed experimentally [see Fig.~\ref{Fig:diffraction2}(b and c)].}
\label{Fig:exclMagStr}
\end{figure*}

We emphasize that the magnetic structure shown in Fig.~\ref{Fig:ModelCalc}(b) of the main text with magnetic space group $P_Ac$ \cite{bilbao} is not unique with respect to the neutron-diffraction data. Systematically searching through the Symmetry-Based Computational Tools for Magnetic Crystallography \cite{bilbao} (Fig.~\ref{Fig:msgGroups} in the main text), we find a few more magnetic structures shown in Fig.~\ref{Fig:inclMagStr} that are consistent with the peak positions in the diffraction measurements. The Bilbao crystallographic database allows for other magnetic structures with the propagation vector of (1/6, 1/6, 0), as shown in Fig. \ref{Fig:exclMagStr}; however, the intensity calculations are inconsistent with the experimental observations. In particular, these configurations show intensities at ($\pm$1/6, $\pm$1/6, 0), which are not observed experimentally. We note that when constructing the magnetic structure using the Bilbao database, we assume that the spin direction at the origin is a high-symmetry direction, and rotate nearest-neighbor moments in the same sublattice by $60^\circ$. For simplicity, we divide the rotation between the spin at the origin and the edge by six, i.e. $360^\circ/6$.


\end{document}